\documentstyle[eqsecnum,preprint,aps,tighten,psfig]{revtex}

\begin{document}
\draft

\title{QCD sum rule analysis for light vector and axial-vector mesons \\ in vacuum
and nuclear matter\thanks{Work supported by GSI Darmstadt and BMBF.}}

\author{Stefan Leupold}

\address{Institut f\"ur Theoretische Physik, Universit\"at
Giessen, D-35392 Giessen, Germany}

\maketitle

\begin{abstract}
Extending previous work we study the constraints of QCD sum rules on mass and width
of light vector and axial-vector mesons in vacuum and in a medium with finite nuclear 
density. For the latter case especially the effect of nuclear pions leading to 
vector--axial-vector mixing is included in the analysis.
\end{abstract}
\pacs{PACS numbers: 14.40.Cs, 21.65.+f, 11.30.Rd, 24.85.+p \\
Keywords: QCD sum rules, meson properties, nuclear matter, chiral symmetry, 
vector--axial-vector mixing}

\section{Introduction}

One of the main goals of modern nuclear physics is to study the behavior of nuclear
matter under extreme conditions. At low temperatures and densities the quarks and gluons
as the basic constituents of strongly interacting matter form hadrons due to the
confinement mechanism. In addition, the appearance of rather light mesons 
(pions and kaons) signals the existence of a spontaneously broken symmetry, 
the chiral symmetry. In fact this symmetry is approximately realized
in the QCD Lagrangian. Another important hint that chiral symmetry is spontaneously
broken in the vacuum state is the absence of chiral partners with equal masses.
In a chirally symmetric state chiral partners would have the same mass.
This concerns for example the isovector vector meson $\rho$ and its much heavier
partner, the isovector axial-vector meson $a_1$. 
It is expected that at high enough temperatures
and densities confinement is lifted and chiral symmetry restored. High energy heavy-ion
collisions are dedicated to the creation of this new state of matter, the 
quark-gluon plasma (QGP) \cite{QM99}. Unfortunately, 
even if such an ultra-hot system of quarks and gluons is created only its decay 
products --- which of course are hadrons and not deconfined quarks and gluons --- 
can reach the detectors. Thus, the proof for the existence of this new stage of
matter has to be performed rather indirectly. 
Besides the observable hadrons also photons and dileptons are radiated
from the hot fireball. These particles deserve special attention since they do not
suffer from strong final state interactions. Therefore, once created in the high
density region they are capable to carry information from that region to the detectors.
Altogether, the challenge is to find unambiguous signs that (part of)
the observed spectra of hadrons, photons, and dileptons are caused by the transient 
existence of the QGP. 
Clearly, to prove the existence of the QGP it is necessary to show
that the spectra cannot be explained by a hot fireball made out of conventional 
interacting hadrons. This task is especially complicated by the fact that there is no
straightforward derivation of hadronic Lagrangians from QCD as the underlying theory
of strong interactions. Therefore it is not a priori clear how far one can trust 
in-medium calculations with hadronic Lagrangians as their parameters are adjusted to
the description of vacuum processes. Connections between hadronic
models and concepts on the one hand side and QCD or QCD based models on the other hand
side are therefore very welcome. The QCD sum rule method \cite{shif79,rubin} provides 
such a link. Before we sketch its basic concepts, however, we want to dwell for a moment
on the in-medium properties of hadrons.

For temperatures and densities below but near to the critical values which mark the 
transition to the QGP it is plausible to expect that already there the properties of
the involved hadrons like e.g.~their masses and decay widths get modified. Especially
the aspect of chiral symmetry restoration is interesting here. The properties of
chiral partners should start to approach each other and finally become identical in the
chirally symmetric phase. Concerning $\rho$ and $a_1$ mesons
possible scenarios are e.g.~discussed in \cite{dey,kapshur,pisarski} for the case of
finite temperature. In principle
one can distinguish three types of possible phenomena (which do not exclude each other):
\begin{itemize}
\item [a)] Mass shifts: The masses of $\rho$ and $a_1$ might approach each other.
One has to distinguish in which way this actually happens: The masses might 
meet at a value
somewhere in between their vacuum masses (and possibly drop together afterwards). It
is however also possible that the masses of both mesons
drop and finally (approximately) vanish at the point of chiral symmetry restoration.
\item [b)] Peak broadening: From the experimental point of view the $\rho$ ($a_1$) meson
shows up as a peak in the vector (axial-vector) channel. In a medium the peaks might 
get broader (maybe without a change of the respective peak positions, i.e.~the nominal 
masses) until the melted spectra in both channels become degenerate. 
\item [c)] Mixing: The distinct peaks might maintain (maybe without shifts or 
broadening), but the $a_1$ peak shows up with increasing height in the vector channel 
and vice versa. 
\end{itemize}
In any case, the spectra in the vector and axial-vector channel become degenerate
when chiral symmetry becomes restored.

In fact, the $\rho$ meson is supposed to be a good candidate to search for a sign
of chiral symmetry restoration. The reason is that it has the quantum numbers of the
photon. Therefore the $\rho$ meson can decay via a virtual photon into a dilepton
pair. If this decay happens within a hot and dense medium the dileptons contain 
information about the in-medium properties of the $\rho$ meson. Therefore, in principle
the possible scenarios discussed above or a mixture of them should leave their marks
in the dilepton spectra. Indeed, the HELIOS and CERES collaborations have reported medium
modifications in the dilepton spectra in the invariant mass range around the $\rho$ meson
mass \cite{CERES,HELIOS}. Whether the observed spectra can be explained within a
conventional hadronic scenario \cite{koch} or whether one has to include medium 
modifications induced by chiral symmetry restoration \cite{brownrho,brownli} is still
a matter of discussion (cf.~also \cite{cassbratrep,rappwam} and references therein). 
Also for the study of 
possible in-medium changes of hadronic properties a closer connection between hadrons
and QCD is desirable. 

The QCD sum rule approach has the merit to relate certain low 
energy quantities --- which so far are not directly accessible by QCD --- with high
energy expressions which can be calculated by the operator product expansion 
\cite{wilson69} in terms of quark and gluon degrees of freedom. Non-perturbative
effects are encoded in the appearance of various quark and gluon condensates.
In the following this method is applied to vector and 
axial-vector mesons placed in a cold medium with finite nuclear density. To clearly
work out the modifications when changing from vacuum to a medium we also discuss
the vacuum sum rule analysis for $\rho$ and $a_1$ in some detail. Before sketching
the basic ideas of the sum rule approach we review the present status of in-medium
analyses for vector and axial-vector mesons especially in the light of the
possible scenarios of in-medium changes discussed above: Concerning finite temperature 
$T$ it has been shown \cite{dey} that at $O(T^2)$ and neglecting the pion mass only 
mixing occurs. A systematic study beyond this linear (pion) density approximation
is complicated by unknown non-scalar higher twist condensates \cite{hats93,ElIo,ElElKa}. 
For finite nucleon density previous analyses have restricted their attention to 
the vector channel. In the first analyses \cite{hats92,hats95} only a possible mass 
shift for the $\rho$ meson has been taken into account, i.e.~the possible scenarios of
peak broadening and/or mixing as mentioned above have been excluded by hand. In this
case it was found that the $\rho$ meson mass would drop in a nuclear medium. However,
it has been shown by the authors of \cite{klingl97} that their specific hadronic model 
also fulfills the sum rule (cf.~also \cite{asakawa1,asakawa2}). 
This model predicts peak broadening for the $\rho$ meson 
and basically no mass shift. Subsequently, a systematic study revealed that independently
of the chosen hadronic model the sum rule for the $\rho$ meson for finite density
is in accordance with a specific mass-width correlation \cite{leupold97}: For low width 
the mass has to decrease. If, however, the mass stays constant --- or even rises ---
the width has to increase. The sum rule does not have enough predictive power to fix 
both the mass and the width of the vector meson. We will come back to that point below.
To the best of our knowledge, the third possible in-medium modification,
the mixing phenomena, has not yet been included in a systematic sum rule analysis for
the vector axial-vector system for finite density. The purpose of the present work is
to treat the properties of $\rho$ and $a_1$ on equal footing, allowing for mass-shifts,
peak broadening, and mixing. 

Before we turn to the specific sum rule analysis for the chiral partners $\rho$ and
$a_1$ we discuss some important aspects of the QCD sum rule approach
focusing especially on in-medium situations.
Recall that our basic motivation was to describe various spectra of heavy-ion collisions
by hadronic models in absence of a QCD description based on first principles. Even when
only hadronic models are capable to calculate {\em observable} 
quantities one can imagine that
it is possible to find other quantities which can be reliably determined both within
the hadronic framework and in terms of quark and gluon degrees of freedom. In this way
one obtains predictions for hadronic parameters like masses and coupling constants
or cross-check for hadronic models. Concerning the QCD sum rule method such 
quantities are specific correlators (see below) calculated in the deep space-like
region, i.e.~for large momenta $q$ with $q^2 \ll 0$. For very large $Q^2 = -q^2$ 
QCD perturbation theory becomes applicable. Proceeding to (somewhat) smaller values
of $Q^2$ non-perturbative corrections appear. They can be expanded in a power series
in $1/Q^2$, called the operator product expansion (OPE):
\begin{equation}
  \label{eq:powermoge}
\sum {c_n \over Q^{2n} }  \,.
\end{equation}
In the coefficients $c_n$ the famous quark and gluon condensates enter. One can imagine
the series (\ref{eq:powermoge}) as a separation of the hard (denominator) and soft 
(numerator) scales of the problem (cf.~e.g.~\cite{leupold98} and references therein). 
In the numerator the non-perturbative effects enter. In practice,
only the first few coefficients in (\ref{eq:powermoge}) can be determined. Of course,
this does not matter as long as $Q^2$ is large enough. Thus, the
crucial question is for which values of $Q^2$ one can trust the truncated series.
If we want to learn something about a hadron with mass $m_h$ it turns out that 
$Q^2$ has to be of the order of $m_h^2$. To get an order of magnitude estimate for
the coefficients $c_n$ we have to ask about the typical scales for non-perturbative 
effects. Let us discuss step by step the different cases of vacuum, finite temperature
and finite baryon density. For vacuum the typical scales are $\Lambda_{\rm QCD}$ and
the current quark masses. The up and down quark masses have only a few MeV and are 
therefore negligibly small. The strange quark mass and $\Lambda_{\rm QCD}$ are between
100 and $200\,$MeV. On the other hand, the typical hadron masses are of the order of
$1\,$GeV. Therefore, one might expect that the sum rule analysis leads to reasonable
results.\footnote{We restrict our considerations here to hadrons made
out of light quarks. The masses of the heavy quarks have to be regarded as part of
the hard scale \cite{shif79}.} 
Of course, the masses of the much lighter pions and kaons cannot
be determined. Unfortunately this optimistic picture is not completely true. In fact, 
there might be non-perturbative effects which introduce an additional hard scale, like
e.g.~instantons \cite{schaefshur}. In this case the series (\ref{eq:powermoge})
would break down for the interesting values of $Q^2$. It seems, however, that the
influence of such effects on the $\rho$ and $a_1$ sum rules is not important.
We therefore {\em assume} throughout this work that the OPE works for the vector
and axial-vector channel. 
Nonetheless, this consideration shows that at present the QCD sum rule approach cannot 
be directly justified from QCD without any additional assumptions. Therefore it should
merely be regarded as a QCD based model and not as QCD itself. Turning to the case
of finite temperature involves new scales. It is common practice to approximate the
low\footnote{At high temperatures it is not reasonable to deal with hadronic degrees of
freedom. If one wants to learn something about hadrons low temperature expansions are
appropriate.} temperature medium by a pion gas. Therefore the new scales are the 
temperature
and the pion mass. Also these quantities are of the order of $\Lambda_{\rm QCD}$.
Of course, $T$ might also be lower. Therefore the previous considerations apply also
here. The case of finite baryon density $\rho_N$ is more complicated. Here one 
approximates
the medium by a Fermi gas of nucleons. New in-medium scales are the Fermi momentum and
the nucleon mass. While the former is reasonably small e.g.~for saturation density of 
nuclear matter, the latter is of the order of $1\,$GeV. Since the nucleon mass enters
the series (\ref{eq:powermoge}) in the numerator it becomes questionable whether the 
OPE still works (cf.~also the discussion in \cite{elio97} and the successive comments 
\cite{reply1,reply2}). Full clarification of this question requires the determination
of all coefficients $c_n$ which would be equivalent to solving QCD in the 
non-perturbative low energy domain. This is of course out of reach. 
For our case at hand there is, however, a class of contributions to the OPE which can
be determined to all orders, namely the twist-two condensates
\cite{hats92,hats95,leupold98,flk99}. In fact their contribution to the coefficients
in the low density approximation is given by
\begin{equation}
  \label{eq:tw2concoeff}
c_n^{\rm twist-two} = a_n m_N^{2n-3} \rho_N   \,,
\end{equation}
i.e.~powers of $Q^2$ in the denominator which in an optimal situation should suppress 
higher order contributions
in (\ref{eq:powermoge}) are compensated by powers of $m_N^2$ in the numerator. Thus
the class of twist-two contributions shows exactly the unpleasant feature discussed
above. In (\ref{eq:tw2concoeff}) the dimensionless quantities $a_n$ can be determined 
from the parton distributions in a nucleon \cite{hats92,hats95,leupold98,flk99}. 
Fortunately
it turns out that $a_n$ is strongly decreasing with increasing $n$ such that the
higher dimensional contributions of the twist-two condensates can safely be neglected
\cite{leupold98,flk99}. This is a hint that the OPE still works in the case of finite
nuclear density. Of course this is not a proof for the validity of the OPE.
Throughout this work we {\em assume} that the OPE works. In spite of these obvious
problems inherent to the QCD sum rule approach for finite density we regard the analysis
presented in the following as useful in view of the possibility
to learn something about the in-medium properties of hadrons from an approach which
deals with the fundamental degrees of freedom of QCD. Nonetheless we stress again
that the QCD sum rule approach --- especially for the case of finite nuclear density --- 
is not as fundamental as QCD.

The article is organized in the following way: In the next section we derive the
in-medium Borel sum rules for $\rho$ and $a_1$ which we will use throughout this work 
for any quantitative statements. In Sec.~\ref{sec:fesr} we make a detour to discuss
a different type of sum rule, namely the finite energy sum rule. This will yield a
qualitative picture what one has to expect from a sum rule analysis and how much
information one can get. In Sec.~\ref{sec:brwi} we introduce our hadronic 
parametrizations which are used to analyze the sum rules. Results for $\rho$ and
$a_1$ in vacuum are presented in Sec.~\ref{sec:resvac}. These results serve as a
reference frame with which we can compare the succeeding in-medium results. 
Sec.~\ref{sec:inmed} is devoted to the discussion of the mixing phenomena while the
in-medium results are presented in Sec.~\ref{sec:resmed}. Finally we summarize and
discuss our results in Sec.~\ref{sec:sum}.

\section{The current-current correlator and the Borel sum rule}  \label{sec:ccbs}

The relevant quantity to look at is the covariant 
time ordered current-current correlator 
\begin{equation}
\Pi_{\mu\nu}(q) = i \int\!\! d^4\!x \, e^{iqx} 
\langle T j_\mu(x) j_\nu(0) \rangle    \,.
  \label{eq:curcur}
\end{equation}
For the $\rho$ meson channel $j_\mu$ is the isospin-1 part of the electromagnetic 
current,
\begin{equation}
\label{eq:curud}
j^V_\mu = {1\over 2} \left( \bar u \gamma_\mu u - \bar d \gamma_\mu d \right)   \,.
\end{equation}
This current-current 
correlator enters e.g.~the cross section of $e^+ e^- \to $ hadrons (see below). 
For the $a_1$ meson channel we have to deal with the corresponding axial-vector current
\begin{equation}
\label{eq:axcurud}
j^A_\mu = {1\over 2} \left( \bar u \gamma_\mu \gamma_5 u 
- \bar d \gamma_\mu \gamma_5 d \right)   \,.
\end{equation}

The expectation 
value in (\ref{eq:curcur}) is taken with respect to the surrounding environment. 
We study here, first, vacuum and, second, an (isospin neutral) equilibrated 
homogeneous medium with finite 
nuclear density and vanishing temperature. 
In the medium Lorentz invariance is broken. All the formulae which we
will present in the following refer to the Lorentz frame where the medium is at
rest, i.e.~where the spatial components of the baryonic current vanish. 
For simplicity we restrict our considerations to mesons which are at rest with 
respect to the medium. 
For the vacuum case we can choose the rest system of the (axial-) vector meson without 
any loss of generality.

In the following the 
formulae without an explicit $V$ or $A$ index are valid for both vector and axial-vector
channel. The correlator (\ref{eq:curcur}) has the following decomposition
(valid for mesons at rest!)
\begin{equation}
  \label{eq:decompmunu}
\Pi_{\mu\nu}(q) = q_\mu q_\nu R(q^2) - g_{\mu\nu} \Pi^{\rm isotr}(q^2)    \,.
\end{equation}
In the following we concentrate on $R(q^2)$. In the vector channel one has
$\Pi^{\rm isotr}(q^2) = q^2 R(q^2)$ since the current $j_\mu^V$ is conserved. We prefer
the use of $R$ instead of $\Pi^{\rm isotr}$ since it has been shown in 
\cite{hats95,leinw,jinlein} that the Borel sum rule (see below) is rather unstable
for the latter quantity. The divergence of the axial-vector channel is solely determined
by the pion decay. Hence we would not learn anything new about the $a_1$ by studying
$\Pi^{\rm isotr}$ in addition to $R$.

Concerning e.g.~the dilepton production one is interested in the values of the 
dimensionless quantity $R_V(q^2)$ in the time-like region $q^2 > 0$. The reason is
that $R_V$ is related to the cross section $e^+ e^- \to $ hadrons with isospin 1 via
\cite{pastar}
\begin{equation}
  \label{eq:crosssec}
{ \sigma^{I=1}(e^+ e^- \to \mbox{hadrons}) \over \sigma(e^+ e^- \to \mu^+ \mu^-) }
= 12 \pi {\rm Im}R_V   \,.
\end{equation}
At least for low energies the time-like region is determined by hadronic degrees 
of freedom. In principle, there are two possibilities to describe 
the current-current correlator. First, guided by an educated 
guess one might use a simple parametrization with some free parameters. Second,
one might use a hadronic model, e.g.~for vector mesons 
\cite{rappwam,klingl97,asakawa1,chanfray92,herrmann,rapp1,friman,rapp2,peters,poust}
using one or the other form of vector meson dominance. In the following we will explore the first 
possibility and figure out which constraints for these free parameters are 
provided by the QCD sum rule approach. For the $a_1$ we proceed completely analogously.
We denote the result for $R$ in the 
time-like region by $R^{{\rm HAD}}$. On the other hand, the current-current 
correlator (\ref{eq:curcur}) can be calculated for $q^2 \ll 0$ 
using Wilson's operator product expansion (OPE) \cite{wilson69} 
for quark and gluonic degrees of freedom 
\cite{shif79} (for in-medium calculations see e.g.~\cite{hats92,hats95}). 
In the following we shall call the result 
of that calculation $R^{{\rm OPE}}$. 
A second representation in the space-like region which has to match 
$R^{{\rm OPE}}$ can be obtained from $R^{{\rm HAD}}$ by
utilizing a subtracted dispersion relation. We find 
\begin{eqnarray}
R^{{\rm OPE}}(Q^2) & = & {\tilde c_1 \over Q^2} + \tilde c_2 
-{Q^2 \over \pi} \int\limits_0^\infty \!\! ds \,
{{\rm Im}R^{{\rm HAD}}(s) \over (s+Q^2)s}
  \label{eq:opehadr}
\end{eqnarray}
with $Q^2:= -q^2 \gg 0$ and some subtraction constants $\tilde c_i$.

Eq.~(\ref{eq:opehadr}) connects hadronic with quark-gluon based expressions.
In principle, for a given hadronic parametrization of $R^{{\rm HAD}}$ with free
parameters this equation could be used to extract information about these 
parameters. This, however, would require the 
knowledge of $R^{{\rm HAD}}(s)$ for arbitrary large $s$. In practice, 
the situation is such that one has a parametrization for the current-current 
correlator for the energy region of the lowest hadronic resonance, but 
one usually has no model which remains valid for arbitrary high energies. 
In the dispersion integral of (\ref{eq:opehadr}) higher lying resonances are 
suppressed, but only by a factor $1/s^2$. Clearly, it is desirable to achieve a
larger suppression of the part of the hadronic spectral distribution on which one has 
less access. To this aim, a Borel transformation \cite{shif79,pastar} 
can be applied to (\ref{eq:opehadr}). For an arbitrary function $f(Q^2)$ the Borel 
transformation is defined as
\begin{equation}
  \label{eq:fftildebo}
f(Q^2) \stackrel{\hat B}{\to} \tilde f(M^2)
\end{equation}
with 
\begin{equation}
  \label{eq:boreldef}
\hat B := \lim\limits_{{Q^2 \to \infty \,, \, N \to \infty 
                         \atop Q^2/N =: M^2 = {\rm fixed} }}
{1 \over \Gamma(N) } (-Q^2)^N \left( {d \over d Q^2} \right)^N   
\end{equation}
where $M$ is the so-called Borel mass. 
Applying the Borel transformation to (\ref{eq:opehadr}) we finally get 
\cite{leupold98}
\begin{eqnarray}
\tilde R^{{\rm OPE}}(M^2)  & = & {\tilde c_1 \over M^2} +
 {1 \over \pi M^2} \int\limits_{0^+}^\infty \!\! ds \,
{\rm Im}R^{{\rm HAD}}(s) 
\, e^{-s/M^2}  \,.
  \label{eq:sumrule}
\end{eqnarray}
We observe that higher resonance states are now exponentially suppressed. 
Note that the subtraction constant $\tilde c_2$ of (\ref{eq:opehadr}) has dropped out. 
The other one, $\tilde c_1$, vanishes in vacuum. In a nuclear medium for a meson at rest, 
we incorporate the Landau damping term in the subtraction constant $\tilde c_1$. This term 
comes 
from the absorption of a space-like meson by an on-shell nucleon. Having incorporated this 
term in $\tilde c_1$ we avoid double counting by restricting the integration 
in (\ref{eq:sumrule}) to the time-like region. 
For a detailed discussion of that point cf.~\cite{florbron}. One gets in the linear 
density approximation
\begin{equation}
  \label{eq:landau}
\tilde c_1 = {\rho_N \over 4 m_N}   \,.
\end{equation}
Eq.~(\ref{eq:sumrule}) is the QCD sum rule which we will utilize in the following. 

Having achieved a reasonable suppression of the energy region above the lowest
lying resonance the integral in (\ref{eq:sumrule}) is no longer sensitive to the
details of the hadronic spectral distribution in that region. For high energies
the quark structure of the current-current correlator is resolved. QCD perturbation
theory becomes applicable yielding
\begin{equation}
  \label{eq:qcdperthe}
{\rm Im}R^{{\rm HAD}}(s) 
=  {1 \over 8 \pi} \left( 1+ {\alpha_s \over \pi} \right)   \qquad 
\mbox{for large $s$.}
\end{equation}
These considerations suggest the ansatz
\begin{eqnarray}
{\rm Im}R^{{\rm HAD}}(s) 
& = &
\Theta(s_0 -s) \, {\rm Im}R^{{\rm RES}}(s) 
+ \Theta(s -s_0) \, {1 \over 8 \pi} \left( 1+ {\alpha_s \over \pi} \right)   
  \label{eq:pihadans}
\end{eqnarray}
where $s_0$ denotes the threshold between the low energy region described by a
spectral function for the lowest lying resonance, ${\rm Im}R^{{\rm RES}}$, 
and the high energy region described by a continuum calculated from perturbative 
QCD. In the following we use $\alpha_s(1\,{\rm GeV}) \approx 0.36$.
Of course, the high energy behavior given in (\ref{eq:pihadans}) is only an 
approximation on the true spectral distribution for the current-current 
correlator. Also the rapid cross-over in (\ref{eq:pihadans}) from the resonance
to the continuum region is not realistic. However, exactly here the suppression
factors discussed above should become effective making a more detailed description
of the cross-over and the high energy region insignificant. 
The price we have to pay for the simple decomposition (\ref{eq:pihadans}) is the
appearance of a new parameter $s_0$, the continuum threshold, which in general 
depends on the nuclear density. We will elaborate
below on the determination of $s_0$. 

To study the content of (\ref{eq:sumrule}) for the vector and axial-vector channel
we need the OPE for the l.h.s. In general, it is given by a Taylor expansion in $1/M^2$:
\begin{equation}
  \label{eq:opegen}
\tilde R^{{\rm OPE}}(M^2) = \sum {c_n \over M^{2n}}   \,.
\end{equation}
In the following we present the formulae for
the case of finite nuclear density $\rho_N$. The vacuum case \cite{shif79} is easily 
obtained by $\rho_N \to 0$. For the vector channel one gets \cite{hats92,hats95}
(for details see also \cite{leupold98} and references therein)
\begin{mathletters}
\label{eq:coeffrho}
\begin{eqnarray}
c^V_0 & = & {1\over 8\pi^2}\left(1+{\alpha_s\over\pi} \right) \,,  \\
c^V_1 & = & 0  \,,  \\
c^V_2 & = & {1\over 24} \left\langle {\alpha_s \over \pi} G^2 \right\rangle 
+ {1 \over 4} m_N A_2 \rho_N
+ m_q \langle \bar q q\rangle    \,,  \\
c^V_3 & = & -{5 \over 24} m_N^3 A_4 \rho_N
-{56 \over 81} \pi\alpha_s \langle {\cal O}^V_4 \rangle    \label{eq:coeff3rho}  
\end{eqnarray}
\end{mathletters}
while for the axial-vector sector one obtains \cite{shif79,hats93}
\begin{mathletters}
\label{eq:coeffa1}
\begin{eqnarray}
c^A_i & = & c^V_i  \quad \mbox{for} \quad i=0,1,2  \,\,,  \\
c^A_3 & = & -{5 \over 24} m_N^3 A_4 \rho_N
+{88 \over 81} \pi\alpha_s \langle {\cal O}^A_4 \rangle   \label{eq:coeff3a1}  \,.
\end{eqnarray}
\end{mathletters}
We neglect (unknown) condensates with dimension higher than 6 and some
less important twist-4 condensates and $\alpha_s$ corrections 
(cf.~\cite{hats95,lee97,leupold98}). We also neglect perturbative contributions proportional
to the square of the current quark masses. 
Note that all expectation values have to be taken 
with respect to the medium. We work here in the linear density approximation:
\begin{equation}
  \label{eq:lindens}
\langle {\cal O} \rangle \approx \langle {\cal O} \rangle_{\rm vac} +
{\rho _N \over 2 m_N} \langle N \vert {\cal O} \vert N \rangle   \,.
\end{equation}
A single nucleon state is denoted by $\vert N \rangle$. It is normalized according to
\begin{equation}
  \label{eq:normnuc}
\langle N(\vec k) \vert N(\vec k') \rangle 
= (2\pi)^3 \, 2 E_k \, \delta(\vec k - \vec k')   \,.
\end{equation}
We defer the calculation of the in-medium expectation values of the scalar operators
to Sec.~\ref{sec:inmed} and only discuss their vacuum expectation values here.
For the gluon condensate we use a canonical value of \cite{shif79}
$\left\langle {\alpha_s \over \pi} G^2 \right\rangle_{\rm vac} = (330 \,\mbox{MeV})^4$.
As compared to the gluon-condensate the influence of the two-quark condensate 
\cite{shif79}
\begin{equation}
  \label{eq:gor}
m_q \langle \bar q q\rangle_{\rm vac}  = - {1 \over 2} f_\pi^2 m_\pi^2
\end{equation}
is rather small (and is further diminished in a nuclear environment). Here 
$f_\pi = 93\,$MeV denotes the pion decay constant and $m_\pi$ the pion mass. While
the values for gluon and two-quark condensate are fairly well known the knowledge
about the four-quark condensates\footnote{Note that the definition of 
$\langle {\cal O}^{V/A}_4 \rangle$ is chosen such that the factorization assumption
would imply $\langle {\cal O}^{V/A}_4 \rangle \approx \langle \bar q q \rangle^2$.}
\begin{equation}
  \label{eq:fourqdef}
\langle {\cal O}^V_4 \rangle =   
{81 \over 224} \left\langle 
(\bar u \gamma_\mu \gamma_5 \lambda^a u - \bar d \gamma_\mu \gamma_5 \lambda^a d)^2 
\right\rangle
+ {9 \over 112} \left\langle 
(\bar u \gamma_\mu \lambda^a u + \bar d \gamma_\mu \lambda^a d)
\sum\limits_{\psi = u, d, s} \bar\psi \gamma^\mu \lambda^a \psi
\right\rangle
\end{equation}
and
\begin{equation}
  \label{eq:fourqdefa1}
\langle {\cal O}^A_4 \rangle =
-{81 \over 352} \left\langle 
(\bar u \gamma_\mu \lambda^a u - \bar d \gamma_\mu \lambda^a d)^2 
\right\rangle
- {9 \over 176} \left\langle 
(\bar u \gamma_\mu \lambda^a u + \bar d \gamma_\mu \lambda^a d)
\sum\limits_{\psi = u, d, s} \bar\psi \gamma^\mu \lambda^a \psi
\right\rangle   \,.
\end{equation}
is very limited.
Traditionally factorization is assumed which, however, probably underestimates its
value. In the following we will use two values for the four-quark condensates
to explore the sensitivity of the results:
\begin{equation}
  \label{eq:fourravac}
\langle {\cal O}^V_4 \rangle_{\rm vac} \, , \langle {\cal O}^A_4 \rangle_{\rm vac}
= (-292 \,\mbox{MeV})^6\, , \, (-281\,\mbox{MeV})^6   \,.
\end{equation}
The larger value is chosen as to obtain an optimal agreement between QCD sum rule
prediction and experiment for the $\rho$ meson properties in vacuum.
Finally the terms proportional to $A_2$ and $A_4$ in 
(\ref{eq:coeffrho},\ref{eq:coeffa1}) stem from twist-2 condensates. They
are obtained from the moments of the quark distributions in a nucleon \cite{hats92}.
We use $A_2 = 0.9$, $A_4 = 0.12$. 

So far we have not specified for which values of $M^2$ we regard the sum rule 
(\ref{eq:sumrule}) to be valid. Note that in practice (\ref{eq:opegen})
is a truncated series 
in $1/M^2$. Clearly, if $M^2$ is too small the $1/M^2$ expansion in
(\ref{eq:opegen}) breaks down. On the other hand, however, 
if $M^2$ is too large the
exponential in (\ref{eq:sumrule}) does not sufficiently suppress the intermediate 
and high energy part of ${\rm Im}R^{{\rm HAD}}(s)$ given in (\ref{eq:pihadans}). As 
mentioned above this suppression is important since the modeling of the region
around the threshold $s_0$ is rather crude. If these qualitative considerations are 
put on a more quantitative level one can define a 
so called Borel window for the masses $M^2$ in which the sum rule is valid 
(cf.~e.g.~\cite{leupold97}). Following \cite{leinw} we determine the minimal Borel mass 
such that the last
accessible contribution to the OPE (\ref{eq:opegen}), i.e.~here the $1/M^6$ term, amounts
to 10\% of the total OPE result:
\begin{equation}
  \label{eq:defmbmin}
\left\vert {c_3 \over M^6_{\rm min}} \right\vert = 
0.1 \tilde R^{\rm OPE}(M^2_{\rm min})  \,.
\end{equation}
The maximal Borel mass is chosen such that the continuum contribution to the r.h.s.~of
(\ref{eq:sumrule}) does not become larger than the contribution from the resonance which 
we want to study, i.e.
\begin{equation}
  \label{eq:mmax}
\int\limits^\infty_0 \!\! ds \,
{1\over 8\pi}\left(1+{\alpha_s\over\pi} \right) \Theta(s-s_0)  \, e^{-s/M^2_{\rm max}} =
\int\limits^\infty_0 \!\! ds \,{\rm Im}R^{{\rm RES}}(s) 
\,\Theta(s_0-s) \, e^{-s/M^2_{\rm max}}  \,.
\end{equation}
As a guideline one can expect that $M^2_{\rm max}$ scales with the point where the average
strength of ${\rm Im}R^{{\rm RES}}(s)$ is located, i.e.~with the resonance mass (squared). 
Hence for large (small) resonance masses the value of $M^2_{\rm max}$ will be large (small).
It might appear that in some cases the Borel window between $M^2_{\rm min}$ and 
$M^2_{\rm max}$ is rather small or even closed. Then
the sum rule is meaningless. In practice the determination of the Borel window
provides a quality check for the sum rule.

\section{Finite energy sum rules --- the qualitative picture}  \label{sec:fesr}

The Borel sum rule (\ref{eq:sumrule}) is not the only sum rule which is used to
connect hadronic and QCD based information. Inserting (\ref{eq:pihadans}) in
(\ref{eq:sumrule}) and expanding the r.h.s.~in powers of $1/M^2$ one can compare
the coefficients of this expansion with the respective ones in the series on the 
l.h.s.~given by (\ref{eq:opegen}). This yields the finite energy sum 
rules\footnote{This derivation is actually oversimplified since it neglects the running
of the coupling constant. For a rigorous derivation cf.~\cite{maltman} and references
therein.}
(presented here for the $\rho$ meson for the vacuum case)
\begin{mathletters}
\label{eq:fesr}
\begin{eqnarray}
\label{eq:fesr1}
{1 \over \pi} \int\limits_0^{s_0} \!\! ds \, {\rm Im}R_V^{{\rm RES}}(s)
- c^V_0 s_0 
& = & 0   \,,
\\  
\label{eq:fesr2}
-{1 \over \pi} \int\limits_0^{s_0} \!\! ds \, s \,{\rm Im}R_V^{{\rm RES}}(s)
+ c^V_0 \, {s_0^2 \over 2} 
& = & c^V_2   \,,
\\
\label{eq:fesr3}
{1 \over \pi} \int\limits_0^{s_0} \!\! ds \, s^2 \,{\rm Im}R_V^{{\rm RES}}(s)
- c^V_0 \, {s_0^3 \over 3} 
& = & 2 c^V_3
\end{eqnarray}
\end{mathletters}
where the coefficients of the OPE are given in (\ref{eq:coeffrho}), evaluated in this 
section for $\rho_N = 0$.

The first two of these sum rules are utilized e.g.~in \cite{weise}. Obviously
the expansion of the r.h.s.~of (\ref{eq:sumrule}) relies on the assumption that
the Borel sum rule obtained by the simple decomposition (\ref{eq:pihadans}) is valid for 
arbitrary high values of $M^2$. As pointed out
above this is doubtful due to the limited knowledge of ${\rm Im}R^{{\rm HAD}}(s)$ 
in the threshold region. Actually the sensitivity of the respective finite energy 
sum rule on the details of ${\rm Im}R^{{\rm HAD}}(s)$ around $s_0$ is increasing 
when going from (\ref{eq:fesr1}) to (\ref{eq:fesr3}). Thus it might be safe to
extract information from the lowest finite energy sum rule(s). Utilizing higher ones,
however, becomes more and more doubtful. This is the reason why we prefer to use
the Borel sum rule. In addition, for the latter one a consistency check on its
validity is provided by the determination of the Borel window.\footnote{A consistency check
for finite energy sum rules might be obtained in the following way: Clearly the 
discontinuity between the resonance and the continuum region in (\protect\ref{eq:pihadans})
is unrealistic. It is only used to avoid new additional parameters. Introducing instead a
smooth cross-over one can test the sensitivity of the finite energy sum rules on these new
parameters which model the cross-over region. If a finite energy sum rule appears to be 
fairly insensitive to these new parameters it might be regarded as useful.}

Nonetheless, the finite energy sum rules can be used to get a qualitative picture
about the connection of the OPE side to resonance parameters like mass and width.
In fact, ${\rm Im}R_V^{{\rm RES}}$ is a mass distribution. Therefore it appears 
natural to define the first two moments of this distribution, i.e.~an average
mass and a width via
\begin{equation}
  \label{eq:avmass}
\bar m ^2 := 
{\int\limits_0^{s_0} \!\! ds \, s \,{\rm Im}R_V^{{\rm RES}}(s) \over
\int\limits_0^{s_0} \!\! ds \, {\rm Im}R_V^{{\rm RES}}(s) }
\end{equation}
and
\begin{equation}
  \label{eq:gausswidth}
\sigma^2 \bar m ^2 := 
{\int\limits_0^{s_0} \!\! ds \, (s-\bar m^2)^2 \,{\rm Im}R_V^{{\rm RES}}(s) \over
\int\limits_0^{s_0} \!\! ds \, {\rm Im}R_V^{{\rm RES}}(s) }    \,.
\end{equation}
Obviously, the finite energy sum rules (\ref{eq:fesr}) can be used to
connect these moments with the condensates (and the continuum threshold):
\begin{equation}
  \label{eq:avmasscond}
\bar m ^2 = {s_0 \over 2} - {c_2 \over c_0 s_0}   \,,
\end{equation}
\begin{equation}
  \label{eq:gawicond}
\sigma^2 = {1 \over \bar m ^2 }
\left( {s_0^2 \over 3} + {2 c_3  \over c_0 s_0} - \bar m ^4  \right)    \,.
\end{equation}
We can learn two things from these simple relations: First, the average mass is
determined by the dimension-4 (gluon and two-quark) condensates and the continuum
threshold while the dimension-6 condensates (here the four-quark condensate) influence
only the width. Second, we do not have enough information at hand to determine all
the phenomenological parameters. In our case at hand we have three of them, namely the 
continuum threshold $s_0$ and the two moments $\bar m^2$ and $\sigma^2$. On the other
hand we only have two equations for these parameters. Traditionally, the use of
QCD sum rules is accompanied by an additional assumption, namely that the width is
negligible. In this case the mass can be determined. 
In general, however, the best we can hope to 
gain are correlations between the free parameters. Being especially interested in
mass and width we can vary $s_0$ and determine the corresponding values for $\bar m$ and
$\sigma$. The result is shown in Fig.~\ref{fig:correl}. (For the four-quark condensate
(\ref{eq:fourravac}) we have chosen the larger value.) The most important thing to
note here is that the width grows with rising mass. 
We will find correlations of this kind again and again throughout this 
work. The qualitative 
understanding of this correlation is obtained from the simple relations 
(\ref{eq:avmasscond},\ref{eq:gawicond}). 
We are reluctant, however, to draw any quantitative conclusions from the previous 
considerations.
In principle we are interested in the properties of the vector and axial-vector
resonances, e.g.~their masses and widths as defined via Breit-Wigner type parametrizations.
In general, these masses and widths are {\em not} identical to the moments $\bar m^2$
and $\sigma^2$ of the distribution ${\rm Im}R^{\rm RES}$. In addition, as outlined
above we doubt the quantitative reliability of the finite energy sum rules (\ref{eq:fesr})
due to their higher sensitivity to the details of the high energy behavior. Hence we prefer
the use of the Borel sum rule (\ref{eq:sumrule}). For a very elaborate use of combinations
of finite energy sum rules we refer to \cite{maltman}.

\section{Breit-Wigner parametrization of the current-current correlator}
\label{sec:brwi}

The only remaining question is how to parametrize 
${\rm Im}R^{{\rm RES}}(s)$ which enters the sum rule (\ref{eq:sumrule}) via
(\ref{eq:pihadans}). Concerning the vector channel, experiments which determine 
e.g.~$e^+ \, e^- \to \pi^+\,\pi^-$ suggest the parametrization
\begin{equation}
  \label{eq:parares}
{\rm Im} R_V^{{\rm RES}}(s) = \pi F_\rho \, {S_\rho(s) \over s}
\end{equation}
Here $F_\rho$ determines the absolute height of the spectrum and $S_\rho$ denotes
the spectral function of the $\rho$ meson which we will specify further below.
Concerning the axial-vector channel not only the $a_1$ but also the pion shows up there.
Hence the parametrization has to be extended to
\begin{equation}
  \label{eq:pararesa1}
{\rm Im} R_A^{{\rm RES}}(s) = \pi F_{a1} \, {S_{a1}(s) \over s} 
+ \pi f_\pi^2 \delta(s - m_\pi^2)   \,.
\end{equation}
The spectral functions are given by
\begin{equation}
  \label{eq:spec}
S(s) = {1 \over \pi} {\sqrt s \,\Gamma(s) \over (s-m^2)^2+s\,\Gamma(s)^2} \,.
\end{equation}
Here $m$ is the mass of the respective meson and $\Gamma$ its width. We stress again
that these Breit-Wigner parameters are not identical to the moments introduced in the
last section; there is only a qualitative correspondence. We denote the
on-shell width by
\begin{equation}
  \label{eq:defonshwidth}
\gamma = \Gamma(m^2)   \,.
\end{equation}
We have to use an $s$-dependent width in (\ref{eq:spec}) for the following reason: 
In the following we will vary $\gamma$ (and other parameters) over large ranges. 
As outlined above the sum rule (\ref{eq:sumrule}) is insensitive to the modeling
of the high energy behavior of ${\rm Im} R^{{\rm RES}}(s)$. In turn, there is a high
sensitivity to the low-energy part. Therefore, especially for large widths
we have to make sure that at threshold the spectral function shows the correct behavior.
On the other hand, we don't want to overweight our parametrizations with too many
independent parameters. Hence, we are aiming at simple parametrizations which reproduce 
correctly the threshold behavior.

In the vacuum the width of the $\rho$ meson is governed by the decay into two pions.
We use the following parametrization:
\begin{equation}
  \label{eq:widthrhovac}
\Gamma_\rho^{\rm decay}(s) = \gamma_\rho {m^2_\rho \over s} 
\left( {p^{\pi\pi}_{\rm rel}(s) \over p^{\pi\pi}_{\rm rel}(m_\rho^2)} \right)^3
\,\Theta(s-4m_\pi^2)
\end{equation}
with the momentum of the pions in the rest frame of the decaying $\rho$ with invariant
mass $\sqrt{s}$:
\begin{equation}
  \label{eq:prelpipi}
p^{\pi\pi}_{\rm rel}(s) = (s - 4 m_\pi^2)^{1/2}/2   \,.
\end{equation}
Concerning the $a_1$ meson in vacuum its width is dominated by the decay into
rho plus pion. For simplicity we neglect the width of the rho meson here and
use
\begin{equation}
  \label{eq:widtha1vac}
\Gamma_{a1}^{\rm decay}(s) = \gamma_{a1} {m^2_{a1} \over s} 
{p^{\pi\rho}_{\rm rel}(s) \over p^{\pi\rho}_{\rm rel}(m_{a1}^2)}
\Theta(s-(m_\rho+m_\pi)^2)
\end{equation}
with the momentum of pion and rho in the rest frame of the decaying $a_1$ with 
invariant mass $\sqrt{s}$:
\begin{equation}
  \label{eq:prelpirho}
p^{\pi\rho}_{\rm rel}(s) = 
\left[(s-(m_\rho+m_\pi)^2)(s-(m_\rho-m_\pi)^2)\right]^{1/2}/(2\sqrt{s})   \,.
\end{equation}
In a nuclear environment a presumably rather sizable collisional width 
(cf.~\cite{poust} and references therein) has to be added to the decay width. The lowest
threshold for a (axial-)vector meson collision with a nucleon is given by the formation
of pion plus nucleon. We assume that the threshold behavior is dominated by the lowest
accessible partial wave. For the $\rho$ meson this is an $s$-wave leading to
\begin{equation}
  \label{eq:widthrhomed}
\Gamma_\rho^{\rm coll}(s) = \gamma_\rho \, 
\left( {1-m_\pi^2 /s \over 1-m_\pi^2/m_\rho^2} \right)^{1/2} \,\Theta(s-m_\pi^2)  \,.
\end{equation}
For the $a_1$ meson it is a $p$-wave:
\begin{equation}
  \label{eq:widtha1med}
\Gamma_{a1}^{\rm coll}(s) = \gamma_{a1} \, {s \over m_{a1}^2}
\left( {1-m_\pi^2 /s \over 1-m_\pi^2/m_{a1}^2} \right)^{3/2} \,\Theta(s-m_\pi^2)  \,.
\end{equation}
For the vacuum case $\Gamma_{\rho/a1}(s)$ in (\ref{eq:spec}) is given by (\ref{eq:widthrhovac})
and (\ref{eq:widtha1vac}), respectively. 
For the case of nuclear medium we restrict ourselves to the two extreme possibilities
that the width is either dominated by decays or by collisions. Hence we explore the
two cases that $\Gamma_{\rho/a1}(s)$ is either given by 
(\ref{eq:widthrhovac})/(\ref{eq:widtha1vac}) or by 
(\ref{eq:widthrhomed})/(\ref{eq:widtha1med}). 

We will treat $F$, $m$, $\gamma$ and
also the continuum threshold $s_0$ (cf.~(\ref{eq:pihadans})) as free parameters.
The aim is to find out how the sum rule (\ref{eq:sumrule}) constrains these
parameters. As a general rule one can at best determine as many parameters of
the hadronic side of the sum rule as one has powers in $1/M^2$ on the OPE 
side \cite{hats95}. 
For the latter we have given in (\ref{eq:coeffrho},\ref{eq:coeffa1}) four orders in 
powers of $1/M^2$. 
However, the perturbative part ($(1/M^2)^0$ part) has already been used to 
determine the high energy behavior in (\ref{eq:pihadans}). Therefore at best 
only three parameters 
of the hadronic spectral distribution can be determined from the sum rule 
(\ref{eq:sumrule}).\footnote{This is also true if the finite energy sum rules were
used instead of the Borel sum rule. The three sum rules 
(\ref{eq:fesr}) provide three constraints on the hadronic spectral 
distribution. See also the discussion in Sec.~\ref{sec:fesr}.} On the other hand, 
we have for each meson four free parameters in the parametrization
(\ref{eq:pihadans},\ref{eq:parares},\ref{eq:pararesa1},\ref{eq:spec}). 
In the traditional sum rule approach \cite{shif79,hats92,hats95,pastar,lee97} 
the width of the 
respective meson resonance is neglected (narrow width approximation). 
In this case the number of free parameters
reduces to three and the sum rule gains predictive power. This however means that
in addition to the QCD input represented by the OPE
one needs further knowledge to extract predictions from QCD sum rules. In vacuum,
this additional input comes from experiments which tell us that e.g.~the $\rho$
meson indeed is a well-defined resonance with a width considerably smaller than
the mass. In contrast, for the in-medium case
it is so far not clear if the pronounced peak structure of the $\rho$ meson 
survives in a nuclear surrounding or if it is washed out 
\cite{klingl97,rapp1,friman,rapp2,peters,poust}
e.g.~by its coupling to resonance-hole states. The $a_1$ meson already has a large
vacuum decay width. In addition, its mass is so high that it is hard to achieve a clear
separation between ${\rm Im}R^{{\rm RES}}$ and the high energy continuum in 
(\ref{eq:pihadans}) \cite{shif79}. We will come back to that point below when discussing
the results of our QCD sum rule analysis for the $a_1$. Like for the $\rho$ meson, medium 
effects in addition presumably lead to an additional broadening of the $a_1$.
To study the influence of the widths of the vector and axial-vector meson on the 
results extracted from QCD sum rules we
will refrain from neglecting $\gamma$ and proceed with our general parametrization
(\ref{eq:spec}). 

\section{Results for vacuum}   \label{sec:resvac}

We shall now explore which values of mass and width of the $\rho$/$a_1$ meson are 
compatible with the sum rule (\ref{eq:sumrule}). For that purpose we
vary the values for $m$ and $\gamma$ in large ranges. For a given pair
of mass and width we tune the remaining
parameters $F$ and $s_0$ such that the agreement between left and right hand side
of (\ref{eq:sumrule}) is best. The resulting minimal deviation $d$ between
l.h.s.~and r.h.s.~is a measure for the compatibility of the chosen mass-width
pair with the sum rule, i.e.~if $d$ is sufficiently small one might conclude
that the chosen pair of mass and width is allowed by QCD sum rules.
We regard the sum rule to be approximately valid in the range given by the Borel window
introduced above. Hence we define the deviation $d$ as an average over this window
(see \cite{leupold97} for further details).

\begin{figure}
\centerline{\psfig{figure=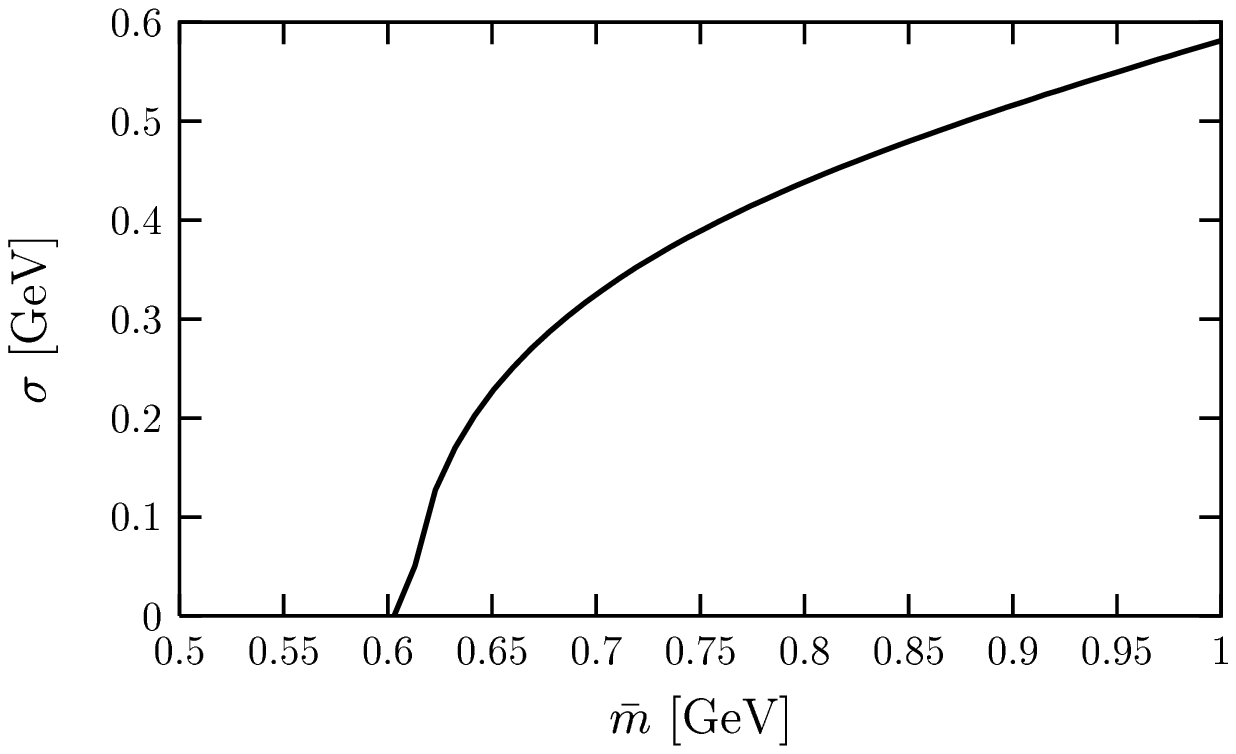,width=12cm}}
\caption{\label{fig:correl} 
Correlation between ``mass'' $\bar m$ and ``width'' $\sigma$ for the $\rho$ meson 
as obtained from the
finite energy sum rules (\protect\ref{eq:avmasscond},\protect\ref{eq:gawicond}). See
main text for details.}   
\end{figure}

\begin{figure}
\centerline{\psfig{figure=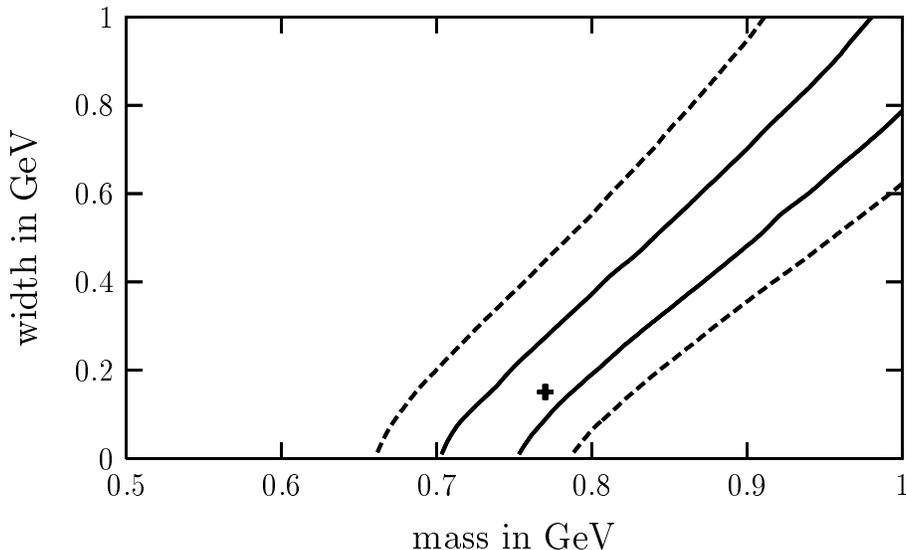,width=12cm}}
\caption{\label{fig:vacrho} 
Deviation $d$ as a function of width and mass 
of the $\rho$ meson for vacuum. For the four-quark condensate a value of 
($-292\,$MeV)$^6$ has been used.
The full lines border the region of QCD sum rule allowed 
parameter pairs with $d \le 0.2\%$, 
the dashed lines border the allowed region 
for $d \le 0.5\%$. The cross marks the experimental values for mass and width
of the $\rho$ meson (including the (very small) error bars according to 
\protect\cite{pdg}).}   
\end{figure}

\begin{figure}
\centerline{\psfig{figure=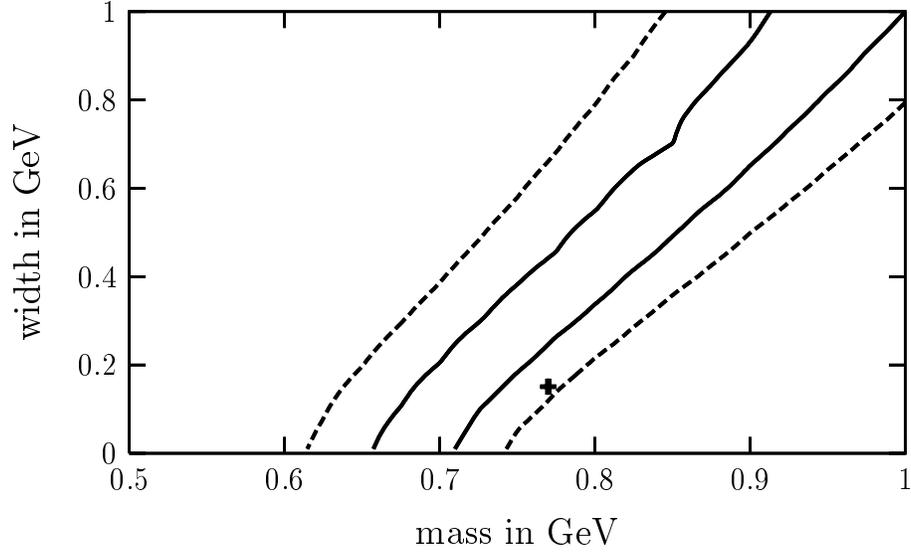,width=12cm}}
\caption{\label{fig:vacrhoold4q} 
Same as Fig.~\protect\ref{fig:vacrho} but with a four-quark condensate of
($-281\,$MeV)$^6$.}   
\end{figure}

For the $\rho$ meson Figs.~\ref{fig:vacrho} and \ref{fig:vacrhoold4q} show the allowed
ranges for mass and width for the vacuum case for the two different values of the
four-quark condensate given in (\ref{eq:fourravac}). 
Obviously, there is not only one point
where the sum rules is reasonably fulfilled but a whole band of allowed mass-width pairs.
Figs.~\ref{fig:vacrho} and \ref{fig:vacrhoold4q} qualitatively resemble
Fig.~\ref{fig:correl}: the band of allowed mass-width pairs describes a correlation where 
the width rises with rising mass. Quantitatively, however, the differences between 
Fig.~\ref{fig:correl} and Figs.~\ref{fig:vacrho}, \ref{fig:vacrhoold4q} are large 
stressing again that the distribution moments defined in
Sec.~\ref{sec:fesr} are not identical to mass and on-shell width of the Breit-Wigner
type spectral functions (\ref{eq:spec}). We have also included the experimental
point for the $\rho$ meson in Figs.~\ref{fig:vacrho} and \ref{fig:vacrhoold4q}. 
Obviously the value of ($-292\,$MeV)$^6$ provides an optimal choice for the four-quark
condensate. A smaller value for that condensate shifts the band of allowed mass-width pairs
either upwards or to the left (or both). From our qualitative considerations of 
Sec.~\ref{sec:fesr} we expect that the four-quark condensate mainly influences the width
(cf.~(\ref{eq:gawicond})). Hence a smaller value of the four-quark condensate is supposed
to increase the width,\footnote{Note that the four-quark condensate enters the sum rule
for the $\rho$ meson with a negative sign, cf.~(\ref{eq:coeff3rho}).} i.e.~to shift the
band upwards. We will further clarify that point when discussing
the results for the $a_1$ meson. Note that also for the smaller value of the four-quark 
condensate of ($-281\,$MeV)$^6$ the deviation $d$ is still reasonably
small for the experimental point (0.46\%).
As discussed above the results of Figs.~\ref{fig:vacrho} and \ref{fig:vacrhoold4q} 
are only meaningful
if the Borel window is reasonably large. For our choice for the condensate values leading
to Fig.~\ref{fig:vacrho} we get
$M^2_{\rm min} = 0.71\,$GeV$^2$. The value for $M^2_{\rm max}$ depends on the resonance
parameters. As already mentioned small values for the resonance mass lead to small values 
for $M^2_{\rm max}$. For the case at hand we find $M^2_{\rm max} > 1.5\,$GeV$^2$ for 
all mass-width pairs lying in the inner band shown in Fig.~\ref{fig:vacrho}
and to the right of it. We regard that as a reasonably large Borel window. 
For Fig.~\ref{fig:vacrhoold4q} the corresponding values are $M^2_{\rm min} = 0.65\,$GeV$^2$
and $M^2_{\rm max} > 1.4\,$GeV$^2$. 

\begin{figure}
\centerline{\psfig{figure=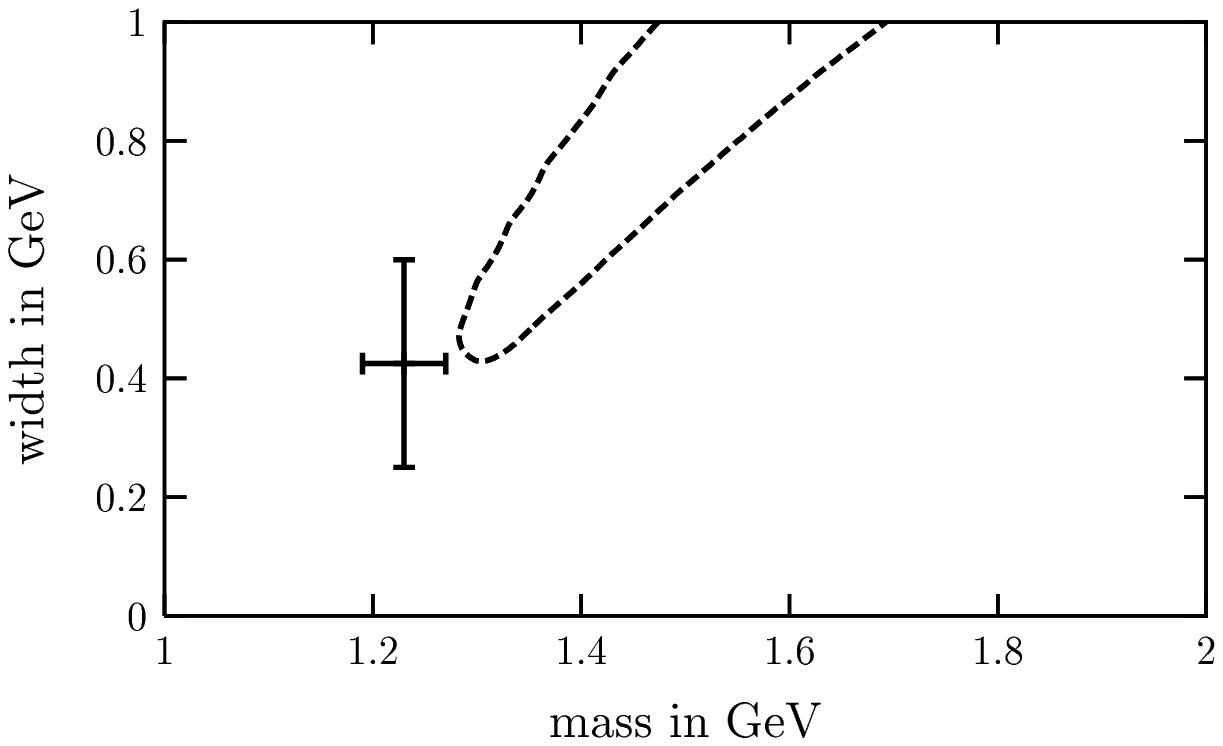,width=12cm}}
\caption{\label{fig:vaca1l4q}
Same as Fig.~\protect\ref{fig:vacrho} for $a_1$ meson in vacuum; value for 
four-quark condensate: ($-292\,$MeV)$^6$.
The cross marks the experimental values for mass and width
of the $a_1$ meson (including the error bars according to 
\protect\cite{pdg}).}   
\end{figure}

\begin{figure}
\centerline{\psfig{figure=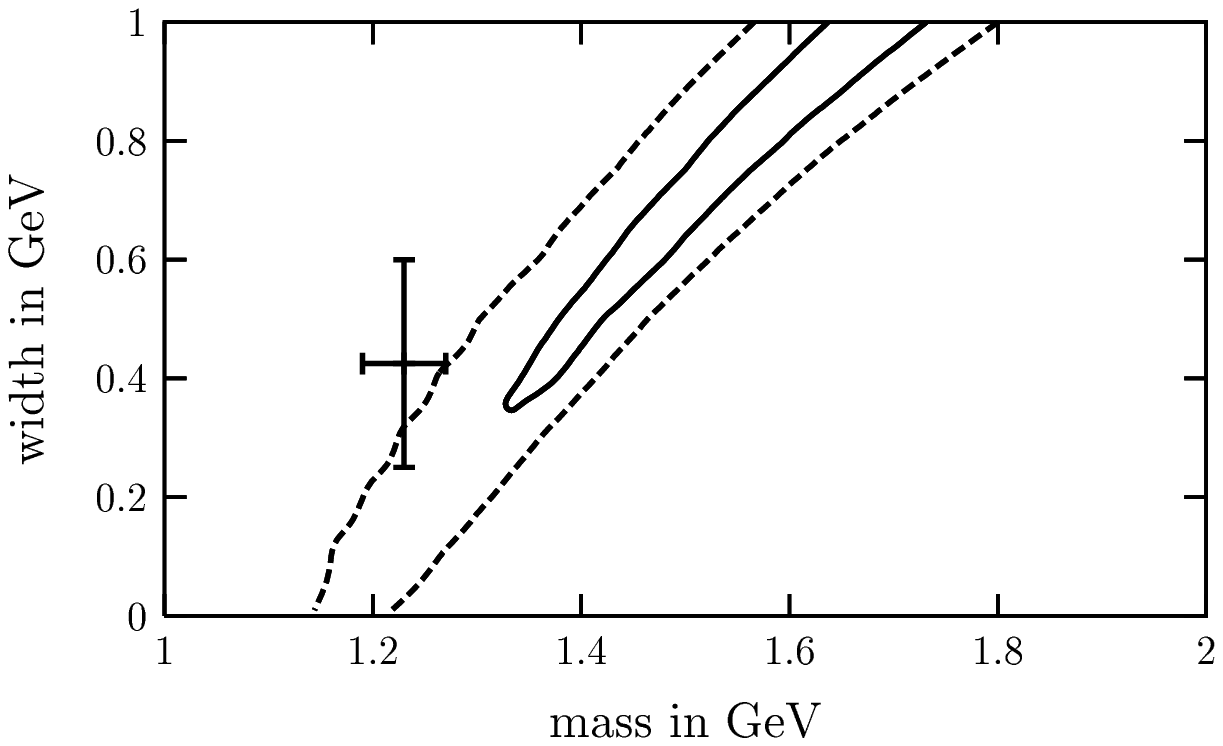,width=12cm}}
\caption{\label{fig:vaca1}
Same as Fig.~\protect\ref{fig:vacrho} for $a_1$ meson in vacuum; value for 
four-quark condensate: ($-281\,$MeV)$^6$.
The cross marks the experimental values for mass and width
of the $a_1$ meson (including the error bars according to 
\protect\cite{pdg}).}   
\end{figure}

Concerning the $a_1$ meson the corresponding mass-width correlations are shown in 
Figs.~\ref{fig:vaca1l4q} and \ref{fig:vaca1}.
Qualitatively we find again the same correlation between masses and widths. However, a
tendency is visible that the sum rule supports large values of the width. We will find that
this tendency increases for the in-medium case. Comparing 
Figs.~\ref{fig:vaca1l4q} and \ref{fig:vaca1} we find that a decrease in the four-quark
condensate shifts the band to some extent to the left but merely downwards. To understand
that finding we note that the four-quark condensate
enters with a different sign in the two sum rules for $\rho$ and $a_1$, respectively;
cf.~(\ref{eq:coeff3rho},\ref{eq:coeff3a1}). Therefore we expect an upward shift with
decreasing four-quark condensate for the $\rho$ meson, as discussed above, and a downward 
shift for the $a_1$. Figs.~\ref{fig:vaca1l4q} and \ref{fig:vaca1} support our
considerations. Comparing the results with the experimental values for mass and width
of the $a_1$ meson we find that the smaller value for the four-quark condensate 
(Fig.~\ref{fig:vaca1}) appears to be much better suited --- quite opposite to the case of 
the $\rho$ meson where the larger value provides a better fit. In principle, there
is no fundamental reason why the four-quark condensates for $\rho$ and $a_1$ should be
exactly the same. Only if the factorization assumption strictly holds the two quantities
defined in (\ref{eq:fourqdef}) and (\ref{eq:fourqdefa1}) coincide. Still, however,
the values ($-292\,$MeV)$^6$ and ($-281\,$MeV)$^6$ are rather close. 
For the succeeding
in-medium calculations we will choose the respective better value, i.e.~($-292\,$MeV)$^6$ 
for the $\rho$ meson and ($-281\,$MeV)$^6$ for the $a_1$. We note that for the $a_1$ vacuum
case a further reduction of the four-quark condensate does not improve the agreement 
between the sum rule and the experimental results as the band is further shifted
downwards and not to the left. Hence the agreement between sum rule and experiment
appears to be better for the $\rho$ than for the $a_1$ meson. This rather old finding
\cite{shif79} is most likely due to the fact that the separation between the resonance
and the continuum region is better realized in the vector channel. If the resonance
appears to be closer to the continuum the sum rule is more sensitive to the details of
the modeling of the nearby transition to the continuum. Such details are necessarily 
rather crude in our ansatz (\ref{eq:pihadans}). Finally we present the results for the
Borel window for the preferable parameter choice (Fig.~\ref{fig:vaca1}):
We find
$M^2_{\rm min} = 0.71\,$GeV$^2$ while all mass-width pairs enclosed by the dashed line
obey $M^2_{\rm max} > 1.4\,$GeV$^2$. We note that we have not included the pion branch
in the determination (\ref{eq:mmax}) of the maximal Borel mass to make sure that we really
learn something about the properties of the $a_1$ meson.

\section{In-medium correlators and $\rho$-$a_1$ mixing}  \label{sec:inmed}

Next we turn to the case of finite nuclear density. As pointed out by several
groups (e.g.~\cite{krippa,chanfray98}) the interaction of the $\rho$ meson with 
the pion cloud of the nucleons induces a mixing of the $\rho$ with its chiral 
partner the $a_1$ meson. This means that e.g.~the (possibly medium-modified) $a_1$-peak
shows up in the spectral distribution ${\rm Im}R^{\rm HAD}_V$ of the {\em vector} 
correlator and vice versa. Suppose now that one would ignore that multi-peak structure 
and still parametrize ${\rm Im}R^{\rm HAD}_V$ with only one peak according to
(\ref{eq:pihadans},\ref{eq:parares}). In a QCD sum rule analysis one has only access on 
certain mass averages of the spectral distribution on account of (\ref{eq:sumrule}).
Hence with a one-peak structure ansatz one would translate certain in-medium changes of 
the OPE side to changes of mass and width of this peak which in reality, however, are 
caused by the appearance of other distinct peaks. This would be rather misleading.
Indeed, for the comparable case of
finite temperature $T$ it has been shown \cite{dey} that the masses of $\rho$ and
$a_1$, if understood as the positions of peaks in spectral distributions, do not change
in the linear (pion) density approximation.\footnote{Note that an $O(T^2)$ modification
at finite temperature corresponds exactly to $O(\rho_N/m_N)$ at finite baryon density
(cf.~e.g.~Eq.~(3) in \cite{dey}).} Only if the notion of mass is used with a different 
meaning (e.g.~in the spirit of Sec.~\ref{sec:fesr} as the first moment of a spectral
distribution) it would be correct to attribute an in-medium $O(T^2)$ mass shift to this 
``mass''. For considerations beyond the $O(T^2)$ approximation we refer to
\cite{ElIo,ElElKa}.

Concerning the present work the mass $m$ of a resonance (which shows up in the 
spectral distribution of the current correlator) is defined via (\ref{eq:spec}).
For small width it gives the peak position of the resonance. In the following for our 
case of finite nucleon density we also try to account for the multi-peak structures
caused by mixing of vector and axial-vector currents. If we only used the sum rule
(\ref{eq:sumrule}) and introduced more than one peak e.g.~in (\ref{eq:parares})
we would have too many free parameters to draw any meaningful conclusion. Therefore
(as for the case of finite temperature \cite{dey}) the key idea is to isolate the
contribution of the respective new in-medium peak(s) also for the OPE side. In this
way one obtains sum rules for non-mixed correlators (in the following called ``bare'')
which can be analyzed with the one-peak ansatz (plus continuum, of course). These
(in general unobservable) bare correlators mix to yield finally the ``full'' in-medium
correlators. The imaginary part of the latter in principle can be observed 
(cf.~e.g.~(\ref{eq:crosssec})).
We note that for the case of finite temperature the bare correlators coincide with
the vacuum correlators in the linear density approximation \cite{dey}. 
As we shall see in the following, things are not so simple for the case of finite 
nucleon density.

To account for the interaction of the nuclear pions with the vector- and axial-vector
currents nuclear matter is separated into a Fermi gas of bare nucleons plus soft 
pions, schematically \cite{krippa}
\begin{equation}
  \label{eq:decnuma}
\vert \Psi \rangle = \Psi^A|A\rangle + \sum_{a}\Psi^A_a |A\pi_a\rangle +
  \sum_{a,b}\Psi^A_{a,b} |A\pi_{a}\pi_{b}\rangle
+ \ldots
\end{equation}
where $\vert \Psi \rangle $ denotes the full nuclear matter state vector while
$\vert A\rangle$ denotes the bare one. The current-current correlator (\ref{eq:curcur}) 
evaluated with respect to $\vert \Psi \rangle $ can now be decomposed into a bare 
correlator, i.e.~a correlator with respect to $\vert A\rangle$, and a part involving the 
interaction with (soft) pions:
\begin{equation}
  \label{eq:pimunubpi}
\Pi_{\mu\nu} = \Pi^b_{\mu\nu} + \Pi^\pi_{\mu\nu} \,.
\end{equation}
The latter one is approximately evaluated using soft pion techniques 
(cf.~also \cite{hats93}) and taking into account up to two pions in the initial 
and/or final state. One gets
\begin{eqnarray}
\langle A\,\pi^a \vert T j_\mu(x) j_\nu(0) \vert A\,\pi^b \rangle
& \simeq &
\langle A\,\pi^a \,\pi^b \vert T j_\mu(x) j_\nu(0) \vert A \rangle
\simeq 
\langle A \vert T j_\mu(x) j_\nu(0) \vert A \,\pi^a \,\pi^b \rangle
\nonumber \\ 
&\simeq  &
{-1\over f_\pi^2}\langle A|\left[Q_5^a ,
\left[Q_5^b , T j_\mu(x) j_\nu(0) \right]\right]|A\rangle   \,,
\label{eq:krsoft}
\end{eqnarray}
with the isovector axial charge
\begin{equation}
  \label{eq:isoaxch}
Q_5^a = \int \!\! d^3\! x \, \bar \psi(x) \gamma_0 \gamma_5 {\tau^a \over 2} \psi(x)  \,.
\end{equation}
To calculate the commutators in (\ref{eq:krsoft}) with the currents (\ref{eq:curud})
and (\ref{eq:axcurud}) it is useful to generalize the latter to the full isospin
multiplets
\begin{equation}
  \label{eq:isomulti}
V_\mu^a = \psi \gamma_\mu \tau^a \psi \quad , \quad 
A_\mu^a = \psi \gamma_\mu \gamma_5 \tau^a \psi   \,.
\end{equation}
In fact, on account of \cite{pastar}
\begin{equation}
  \label{eq:mixcur}
[ Q_5^a , V_\mu^b ] = i \epsilon^{abc} A_\mu^c \quad , \quad 
[ Q_5^a , A_\mu^b ] = i \epsilon^{abc} V_\mu^c \quad , \quad 
\end{equation}
the vector and axial-vector currents are mixed by their interaction with the nuclear 
pions. Finally the expectation values in (\ref{eq:krsoft}) have to be weighted by
the density of pions in the nuclear medium. For the dimensionless quantities defined
in (\ref{eq:decompmunu}) one ends up with (see \cite{krippa} for details)
\begin{mathletters}
  \label{eq:mixing}
\begin{eqnarray}
R_V(q^2) & = & R_V^b(q^2) - \xi \, (R_V^b(q^2) - R_A^b(q^2))  \,  \\
R_A(q^2) & = & R_A^b(q^2) - \xi \, (R_A^b(q^2) - R_V^b(q^2))
\end{eqnarray}
\end{mathletters}
where $R^b_{V/A}$ denotes the respective correlator with respect to a system of 
{\em bare} nucleons, i.e.~without their pionic cloud. The mixing parameter
is given by
\begin{equation}
  \label{eq:defxi}
\xi = {4 \over 3} {\sigma^\pi_N \rho_N \over f_\pi^2 m_\pi^2}   \,.
\end{equation}
Here we have introduced \cite{chanfray98}
\begin{equation}
  \label{eq:defsigmapiN}
\sigma^\pi_N = {m_\pi^2 \over 4 m_N} \langle N \vert \vec \pi^2 \vert N \rangle
= {m_\pi \over 2} N_\pi
\end{equation}
where $N_\pi$ denotes the scalar number of pions in the cloud surrounding the nucleon. 
$\sigma^\pi_N$ contributes to the nucleon sigma term $\sigma_N$ given by \cite{leupold98}
\begin{equation}
  \label{eq:defsigma}
\sigma_N = {m_q \over m_N} \langle N \vert \bar q q \vert N \rangle  \,.
\end{equation}
Having split up the nucleons into bare nucleons plus a cloud of soft pions we
have to disentangle the nucleon sigma term correspondingly \cite{chanfray98}:
\begin{equation}
  \label{eq:splitnuclsig}
\sigma_N = \sigma^b_N + \sigma^\pi_N   \,.
\end{equation}
At present, the value and even the sign for $\sigma^\pi_N$ is not a settled issue.
We take a {\em positive} value of $25\,$MeV. This
is in agreement with \cite{chanfray98} but in contrast to \cite{krippa} where
a negative value has been used. Our choice for $\sigma^\pi_N$ is motivated by the fact
that this ensures that the two
correlators $R_V$ and $R_A$ become degenerate at some finite density:
\begin{equation}
  \label{eq:difdeg}
R_V - R_A = (1 - 2 \xi) (R_V^b - R_A^b)   \,.
\end{equation}
A negative value for $\sigma^\pi_N$ would lead to anti-mixing, i.e.~in this case the
nuclear pions would work against chiral symmetry restoration.

Next we will perform a sum rule analysis for the correlators with respect to
the system of bare nucleons $R^b_{V/A}$
at nuclear saturation density $\rho_N = 0.17/$fm$^3$. Concerning the vector channel 
this is the essential
difference as compared to our previous work \cite{leupold97} where we have
analyzed the sum rule for the full in-medium correlator $R_V$. Note that 
(\ref{eq:coeffrho}),~(\ref{eq:coeffa1}) are valid for both full and bare correlators.
The difference appears in (\ref{eq:lindens}) in the calculation of the expectation 
value with respect to bare or full nucleons, respectively. In practice, 
the difference manifests itself only in the different handling of the four-quark 
condensate.
All OPE contributions in (\ref{eq:coeffrho}) and
(\ref{eq:coeffa1}) except from the two- and four-quark condensates come from chiral 
singlet operators. They do not
distinguish between bare nucleons and nucleons dressed by soft pions. Hence their
evaluation is standard \cite{leupold97}. Things are different for the two- and
four-quark condensates. Concerning the two-quark condensate, its
in-medium change (in linear density approximation) is determined by the nucleon sigma 
term (\ref{eq:defsigma}). To calculate the in-medium change with respect to bare
nucleons we have to take into account only $\sigma^b_N$ as defined via 
(\ref{eq:splitnuclsig}):
\begin{equation}
  \label{eq:scalb2q}
{\langle \bar q q \rangle_b \over \langle \bar q q \rangle_{\rm vac}} 
= 1 - {\sigma^b_N \rho_N \over f_\pi^2 m_\pi^2}   \,.
\end{equation}
In practice, the value of $m_q \langle \bar q q \rangle$ is rather small (as compared
to the gluon condensate) and further diminished in the medium. In contrast, the four-quark
condensate is numerically important. In lack
of a better access to the value of ${\cal O}_4^V$ at finite density we made in 
\cite{leupold97} the common assumption that this four-quark condensate scales with
the density like the square of the two-quark condensate 
\begin{equation}
  \label{eq:fourdens}
\langle {\cal O}_4 \rangle_{\rm med} = 
 \langle {\cal O}_4 \rangle_{\rm vac} 
\left( 
{\langle \bar q q \rangle_{\rm med} \over \langle \bar q q \rangle_{\rm vac}}
\right)^2 =  
\langle {\cal O}_4 \rangle_{\rm vac} 
\left( 1 - {\sigma_N \rho_N \over f_\pi^2 m_\pi^2}  \right)^2  \,.
\end{equation}
In the comparable case
of finite temperature (i.e.~in a hot pion gas), however, it was shown in 
\cite{eletsky} that such an assumption is wrong. This can be traced back to the 
fact that in the presence of pions the two-quark condensate behaves 
different as compared to the four-quark condensate due to its different transformation 
properties with respect to chiral transformations. This suggests that
also at finite nucleon density the scaling assumption (\ref{eq:fourdens}) is 
doubtful due to the presence of virtual pions. In the present work we
have explicitly taken into account the contribution from the pion cloud of the 
nucleons. In this way we have expressed the full correlator in terms of the bare
correlators. We now assume the scaling property (\ref{eq:fourdens}) only for the
condensates with respect to a system of bare nucleons. It takes the form
\begin{equation}
  \label{eq:fourdensbare}
\langle {\cal O}_4 \rangle_b = 
\langle {\cal O}_4 \rangle_{\rm vac} 
\left( 1 - {\sigma^b_N \rho_N \over f_\pi^2 m_\pi^2}  \right)^2  
\end{equation}
where for consistency we have to take the bare nucleon sigma term 
$\sigma^b_N = \sigma_N - \sigma_N^\pi \approx 20\,$MeV instead of the full one 
$\sigma_N \approx 45\,$MeV. In the following we use this scaling assumption 
(\ref{eq:fourdensbare}) for both $\langle {\cal O}^V_4 \rangle_b$ and
$\langle {\cal O}^A_4 \rangle_b$. 

In view of the uncertainty connected with the four-quark condensate it clearly 
would be fortunate to use sum rules which do not involve
it. Indeed, in \cite{weise} the first two finite energy sum rules
(\ref{eq:fesr1},\ref{eq:fesr2}) were used. However, with the same parameter set 
$(m_\rho,\gamma,F,s_0)$ characterizing the hadronic correlator the two sum rules
are capable to determine only two of these four parameters.
To further restrict the parameter space additional information is required. 
In \cite{weise} it is suggested that the threshold $s_0$ is connected to the
scale of chiral symmetry breaking. Thus, the choice is either to make assumptions
about the four-quark condensate or about the threshold parameter. As outlined above we 
prefer to work with the Borel sum rule instead of the finite energy sum rules due to the 
larger sensitivity of the latter to the high energy modeling. In this case we cannot get
rid of the four-quark condensate.

The sum rule analysis proceeds along the same lines as described above for the
vacuum case. We analyze the Borel sum rules for the $\rho$ as well as the 
$a_1$ meson placed in a medium of {\em bare} nucleons. For the time-like part
of the correlator in the vector channel we use again a single resonance parametrization 
of type (\ref{eq:parares}). For the axial-vector channel we recall that there is
a pion branch in addition to the $a_1$.
Here there is an additional change in the medium due to a change of the pion decay 
constant in nuclear matter. We replace $f^2_\pi$ in (\ref{eq:pararesa1}) by 
\begin{equation}
  \label{eq:fpistar}
{f^*_\pi}^2 \approx f^2_\pi 
{\langle \bar q q \rangle_b \over \langle \bar q q \rangle_{\rm vac}}
\end{equation}
where we have utilized the in-medium version of the Gell-Mann--Oakes--Renner relation
(\ref{eq:gor}) and neglected a possible in-medium change of the pion mass. 
By using (\ref{eq:pararesa1}) with the replacement (\ref{eq:fpistar}) we also assume that
the delta-type spectral function of the pion is not significantly smeared out. In fact,
pion properties are expected to change drastically in nuclear matter due to the strong
coupling to Delta-hole states \cite{weiseerc}. However, this $p$-wave coupling is not
important here since we deal with correlators which are at rest with respect to the
nuclear environment. Working in the linear density approximation, i.e.~neglecting Fermi
motion, the pions do not couple to the nucleons by exciting Deltas. In the context of the
(necessary) approximations involved it seems reasonable to work with a pion spectral
function which is neither shifted nor broadened.

\section{Results for nuclear matter}   \label{sec:resmed}

\begin{figure}
\centerline{\psfig{figure=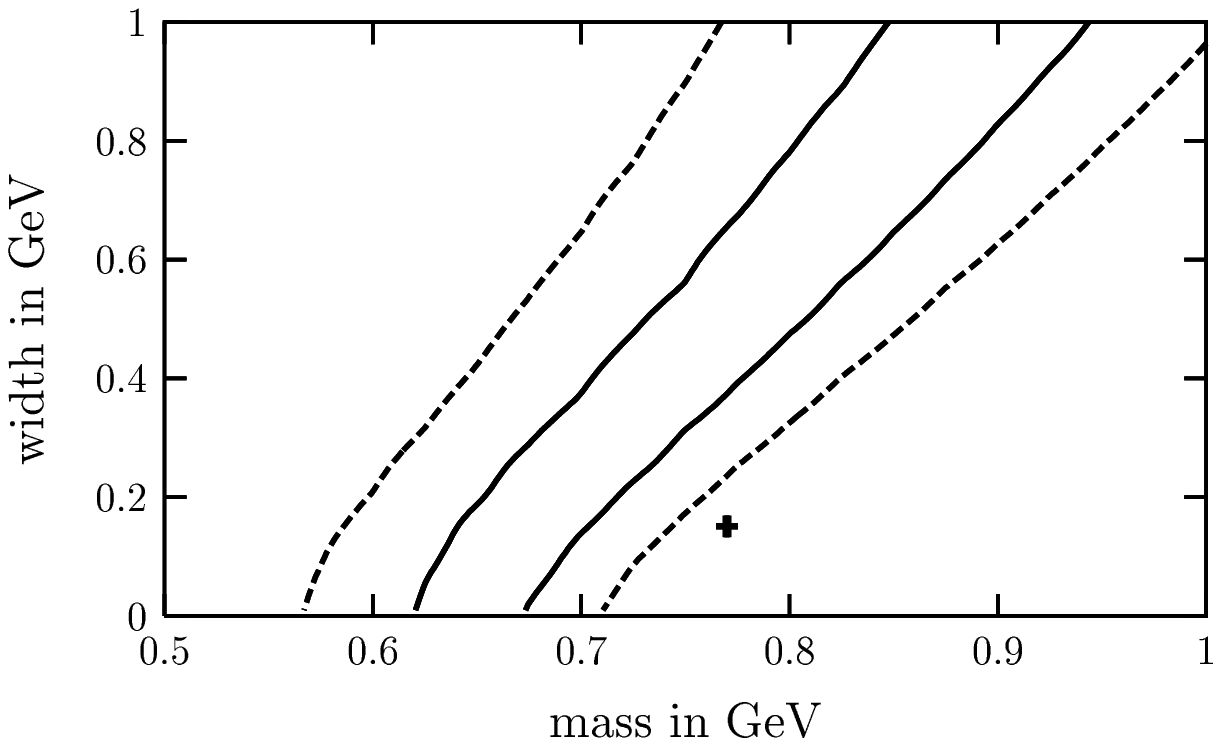,width=12cm}}
\caption{\label{fig:medrho2pi} 
Same as Fig.~\protect\ref{fig:vacrho} for $\rho$ in system of bare nucleons
at normal nuclear matter density. For the calculation of the width the two-pion 
threshold is adopted. See main text for details.}
\end{figure}

\begin{figure}
\centerline{\psfig{figure=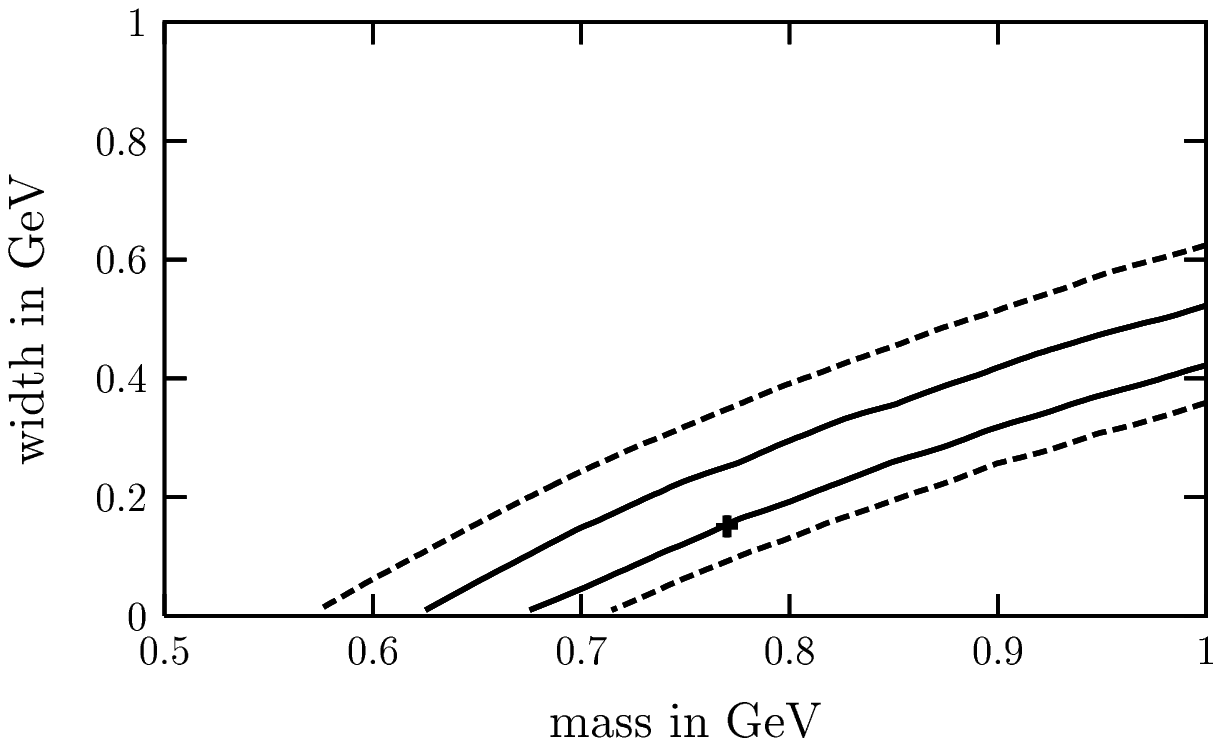,width=12cm}}
\caption{\label{fig:medrho} 
Same as Fig.~\protect\ref{fig:vacrho} for $\rho$ in system of bare nucleons
at normal nuclear matter density. For the calculation of the width the one-pion 
threshold is adopted. See main text for details.}
\end{figure}

\begin{figure}
\centerline{\psfig{figure=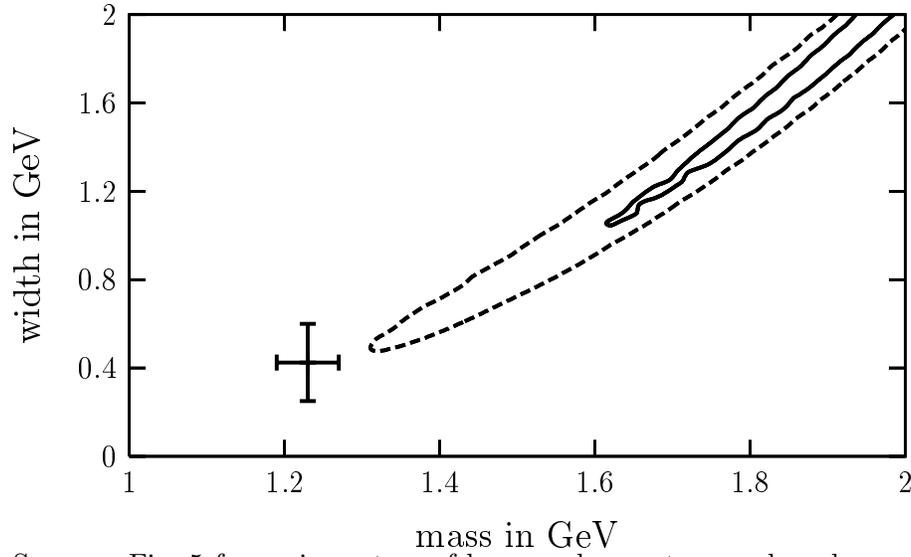,width=12cm}}
\caption{\label{fig:meda1} 
Same as Fig.~\protect\ref{fig:vaca1} for $a_1$ in system of bare nucleons
at normal nuclear matter density. For the calculation of the width the one-pion 
threshold is adopted. See main text for details. Note the different scale for the width
as compared to previous figures.}
\end{figure}

The results of the sum rule analysis are shown in Figs.~\ref{fig:medrho2pi}, 
\ref{fig:medrho}, and \ref{fig:meda1}. As for the vacuum case we find bands of 
allowed mass-width pairs. The interesting point is how these bands have changed as 
compared to the respective vacuum case. As already mentioned concerning the choice for
the vacuum four-quark condensate we have used the respective better value, 
i.e.~($-292\,$MeV)$^6$ for the $\rho$ meson and ($-281\,$MeV)$^6$ for the $a_1$. Hence
we have to compare Figs.~\ref{fig:medrho2pi}, \ref{fig:medrho} to Fig.~\ref{fig:vacrho}
and Fig.~\ref{fig:meda1} to Fig.~\ref{fig:vaca1}. 

For the $\rho$ meson we have explored two
possibilities to parametrize the energy dependence of the width. The same
parametrization (\ref{eq:widthrhovac}) as for the vacuum case, i.e.~with the two-pion
threshold, is used to obtain the results depicted in Fig.~\ref{fig:medrho2pi}. Here 
the mass-width band is shifted to the left as compared to the vacuum case
(Fig.~\ref{fig:vacrho}). Obviously, the in-medium change of the
condensates calls for more strength at lower invariant masses as compared to the situation
in vacuum. This can be accomplished either by a lower peak mass or by a larger width.
For Fig.~\ref{fig:medrho} the parametrization (\ref{eq:widthrhomed}) with the one-pion 
threshold is used. For very small width the masses allowed by the sum rule analysis
agree for Figs.~\ref{fig:medrho2pi} and \ref{fig:medrho}. This is of course due to the
fact that the details of the off-shell parametrization of the width do not matter if
the width is sufficiently small. For larger widths, however, there appear large 
differences between Figs.~\ref{fig:medrho2pi} and \ref{fig:medrho}. The band in 
Fig.~\ref{fig:medrho} is much less steep. This is easy to 
understand: The demand for more strength at lower invariant masses is easier to fulfill
if there is already some strength below the two-pion threshold. Therefore, for the same 
peak mass the on-shell width can be smaller if the one-pion threshold is used instead 
of the two-pion threshold. The large differences between Figs.~\ref{fig:medrho2pi} 
and \ref{fig:medrho} stress again that the Borel sum rule is very sensitive to the
low-energy behavior of the spectral distribution. More generally one has to realize that
peak mass and on-shell width are sufficient to characterize a spectral function only
if the width is not too large. For large width details of the spectral shape become 
important, in the case at hand especially the details in the low-energy region.
Concerning the Borel mass window we get $M^2_{\rm min} = 0.64\,$GeV$^2$. For all
mass-width pairs in the inner bands of Figs.~\ref{fig:medrho2pi} and \ref{fig:medrho}
and to the right of it we find $M^2_{\rm max} > 1.2\,$GeV$^2$. It is a generic
finding of sum rule analyses that the Borel window shrinks to some extent when
changing from the vacuum to the in-medium case \cite{leupold97}. Still we regard the
Borel window to be large enough to draw conclusions from the analysis.

Turning to the $a_1$ meson Fig.~\ref{fig:meda1} shows that the tendency of the sum rule
to support
large values of the width (cf.~Fig.~\ref{fig:vaca1}) increases in 
medium. (Note that 
the width-scale in Fig.~\ref{fig:meda1} differs from the previous figures.) The minimal
Borel mass is given by $M^2_{\rm min} = 0.61\,$GeV$^2$ while 
$M^2_{\rm max} > 2.0\,$GeV$^2$ for the whole relevant mass-width range.
For the width we have used parametrization (\ref{eq:widtha1med}) with the one-pion 
threshold to obtain Fig.~\ref{fig:meda1}. We have also analyzed the sum rule using 
instead (\ref{eq:widtha1vac}) with 
the rho-pion threshold. In this case we did not find any mass-width pair with a 
deviation $d$ less than 1\%. Therefore, we do not show a plot for the latter case.
In fact, even if we allowed for larger values of $d$ there would be 
no sign for the desired mass-width band. We conclude that parametrization
(\ref{eq:widtha1vac}) is incompatible with the in-medium sum rule for the $a_1$ meson. 
Obviously this sum rule demands for a spectral distribution which is smeared out over
a large invariant mass range. Restricting the mass range by a (rather high) lower
limit of $m_\pi + m_\rho \approx 0.9\,$GeV appears to be insufficient to fulfill the
sum rule. Instead, the one-pion threshold (caused by nucleon-$a_1$ collisions) does the 
job. Of course, also other scenarios which might fulfill the sum rule are conceivable.
As we have learned from the previous in-medium sum rule analysis the spectral
function of the $\rho$ meson gains strength at lower invariant mass (either by a lower
peak mass or by a larger width). Thus the effective threshold for the in-medium decay of 
the $a_1$ meson into rho plus pion is probably lowered as compared to the vacuum case.
Such a scenario might also be in line with the sum rule. We have not explored this
possibility in further detail since we would have to introduce a couple of new free
parameters to model e.g.~the successive in-medium decays $a_1 \to \rho + \pi \to
3 \pi$. As already noticed when discussing the results for the $\rho$ meson we find also
for the $a_1$ that the QCD sum rule is sensitive to the threshold modeling of the width 
provided 
that the on-shell width is not too small. This fact has not been sufficiently taken
into account in \cite{leupoldproc} leading to results for the $a_1$ meson which differ 
from the ones presented here. 
We also deduce from Fig.~\ref{fig:meda1} that the nominal peak mass is
shifted to higher values. It is important to note that this shift is {\em not} in 
contradiction to chiral symmetry. What we have studied so far is the behavior of
the correlators under the influence of the system of {\em bare} nucleons. Chiral symmetry
only demands that the {\em full} correlators of the vector and axial-vector channel 
become degenerate at
high enough density. This is achieved by the pions as expressed by the mixing formula 
(\ref{eq:difdeg}) no matter how the $a_1$ mass changes under the influence of the bare 
nucleons. 

To visualize the effect of mixing we now turn to the full in-medium correlators
obtained from the bare ones via (\ref{eq:mixing}). As already 
pointed out the sum rule analysis in general is not capable to pin down both 
in-medium mass and width for the respective meson. Nonetheless, to illustrate
the effect of mixing we arbitrarily take one pair of values for the $\rho$ meson 
from the inner band depicted in Fig.~\ref{fig:medrho} and the optimal pair for the 
$a_1$ meson. We choose: 
\begin{mathletters}
\label{eq:specval}
\begin{eqnarray}
m_\rho = 0.77\,\mbox{GeV,} \quad \gamma_\rho = 0.21\,\mbox{GeV,} & \quad &
s_0^\rho = 1.15\mbox{GeV}^2 , \quad F_\rho = 1.1\cdot 10^{-2} \, \mbox{GeV}^4 \,; \\
m_{a1} = 1.86\,\mbox{GeV,} \quad \gamma_{a1} = 1.67\,\mbox{GeV,} & \quad &
s_0^{a1} = 3.84\mbox{GeV}^2 , \quad F_{a1} = 0.36 \, \mbox{GeV}^4 \,.
\end{eqnarray}
\end{mathletters}
The results for the imaginary part of the 
full in-medium vector and axial-vector correlators are
shown in Figs.~\ref{fig:fullspecrho} and \ref{fig:fullspeca1} 
together with the respective contributions from
the bare correlators. In the figures we have not included the delta function type
contributions from the pion and from Landau damping.
Note that the shoulder of the $\rho$ contribution at low 
invariant mass is caused by
the $1/s$ factor present in (\ref{eq:parares}). Especially the axial-vector correlator
shows that in the medium the resonance peaks are no longer higher than the 
high energy continuum. Recalling that the modeling of the onset of the continuum is
rather crude one might further soften the crossover regions leaving basically no room
for distinct peak structures.
In total, we see a clear sign that the in-medium spectral distributions get washed
out with increasing density.

\section{Summary}   \label{sec:sum}

We have presented a QCD sum rule analysis for the in-medium current-current
correlators with the quantum numbers of $\rho$ and $a_1$ mesons. For the medium we have
chosen the case of a Fermi gas of nucleons at vanishing temperature. For comparison
we have also presented the vacuum sum rule analysis for $\rho$ and $a_1$.
As possible in-medium changes 
for the spectral distributions of both correlators we have allowed for mass-shifts,
peak broadening and also mixing. The latter effect has not been considered in previous
analyses \cite{hats92,hats95,leupold97}. 

In the QCD sum rule only a mass-averaged
quantity involving the respective spectral distribution enters. In general the sum rule
is not capable to pin down the full information present in the spectral
distribution like the number of peaks, their positions, widths and heights. Therefore
one needs an ansatz for the spectral distribution with some free parameters. These
parameters can be constrained by the requirement that the spectral distribution has to
fulfill the sum rule. 
The simplest ansatz consists of only one peak (in the low energy regime) and neglecting
its width. In this case the sum rule for the $\rho$ meson
demands an in-medium mass shift towards smaller masses \cite{hats92,hats95}. At
normal nuclear matter density the mass has to drop by roughly 16\%. This, however,
is a model dependent statement since a specific choice for the form of the
spectral distribution has been made, namely only one peak with vanishing width. 
In fact, if also the width is included but still 
using a one-peak ansatz it is already
impossible to fix both the mass and the width. One obtains a band of allowed mass-width 
pairs \cite{leupold97}. At small width this band of course has to start at the mass 
determined in \cite{hats92,hats95} --- provided the same condensate values are used
(cf.~\cite{leupold97} for details). For larger widths the allowed masses also increase.
With increasing nucleon density the mass-width band is shifted to the left, i.e.~for a 
given value of the width to smaller masses. 

If the true in-medium spectral distribution possesses more than one peak but in the
ansatz for that quantity only one peak was present then certain in-medium changes
caused by the additional peaks would be erroneously attributed to changes of mass
and/or width of the single peak. Therefore, in the present work we have made a further 
step to consider all possible in-medium
modifications. The mixing of the vector with the axial-vector correlator by the
virtual nuclear pions has been included. Using soft pion techniques this mixing
can be approximately calculated for the correlators irrespective of the choice for 
the correlator momentum $q$.
Therefore the correlators can be decomposed into a superposition of ``bare'' correlators
for both the OPE side ($q^2 \ll 0$) and
the spectral distribution ($q^2 > 0$). In this way we have split off the mixing
phenomena (at least the one induced by nuclear pions) from the sum rule analysis. 
Successively the latter has been applied to the bare correlators. 
As compared to previous works we have obtained a less drastic in-medium shift of the 
allowed mass-width band for the $\rho$ branch. For the $a_1$ branch we have found
that the sum rule is better fulfilled for not too small values of the width. The 
preference of large widths increases with increasing density. 
We have also seen that for both
the $\rho$ and $a_1$ meson at larger widths the respective sum rule is sensitive to the 
threshold modeling of the energy dependence of the width. Therefore, it appears to
be insufficient to characterize a peak in the spectral distribution by its position
and on-shell width. A reasonable modeling of the energy dependence of the width is
important to obtain meaningful results from a QCD sum rule analysis. 

\begin{figure}
\centerline{\psfig{figure=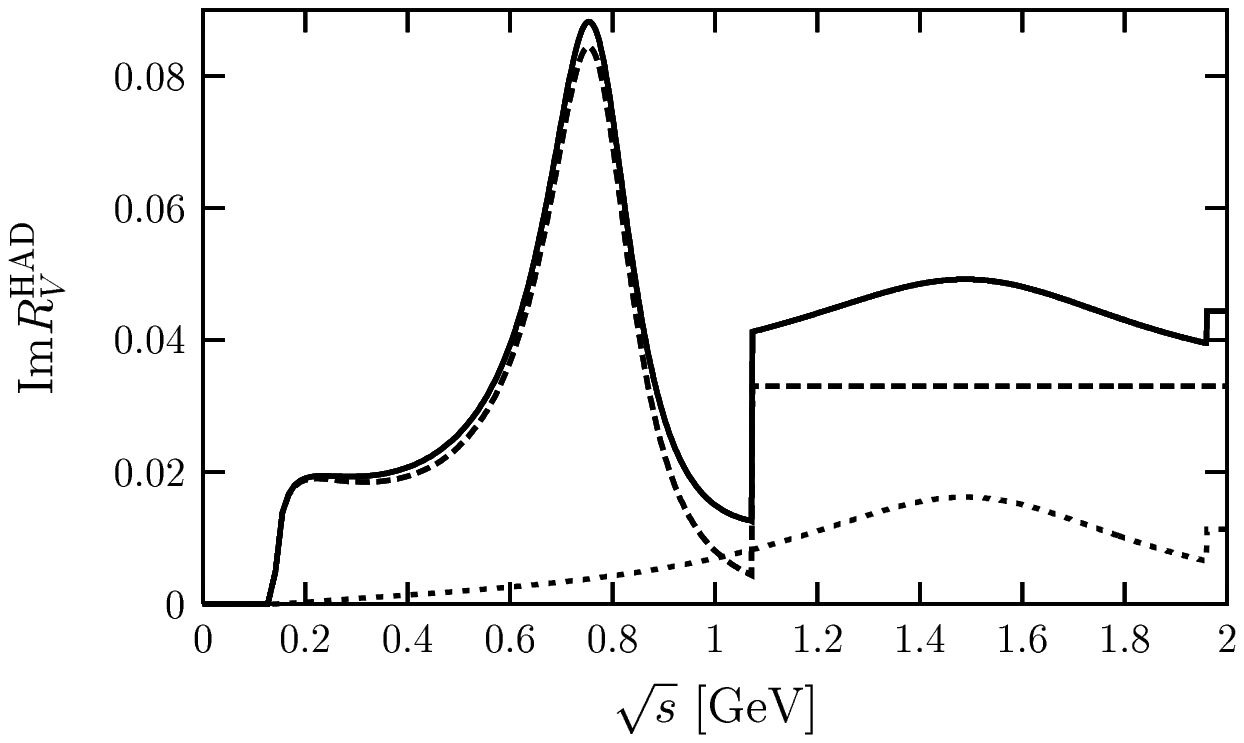}}
\caption{\label{fig:fullspecrho}
Imaginary part of the full in-medium vector correlator (full line) as a 
function of $\sqrt{s} = \sqrt{q^2}$. The contribution of the bare vector (axial-vector)
correlator is the given by the dashed (dotted) line. See main text for details.}
\end{figure}

\begin{figure}
\centerline{\psfig{figure=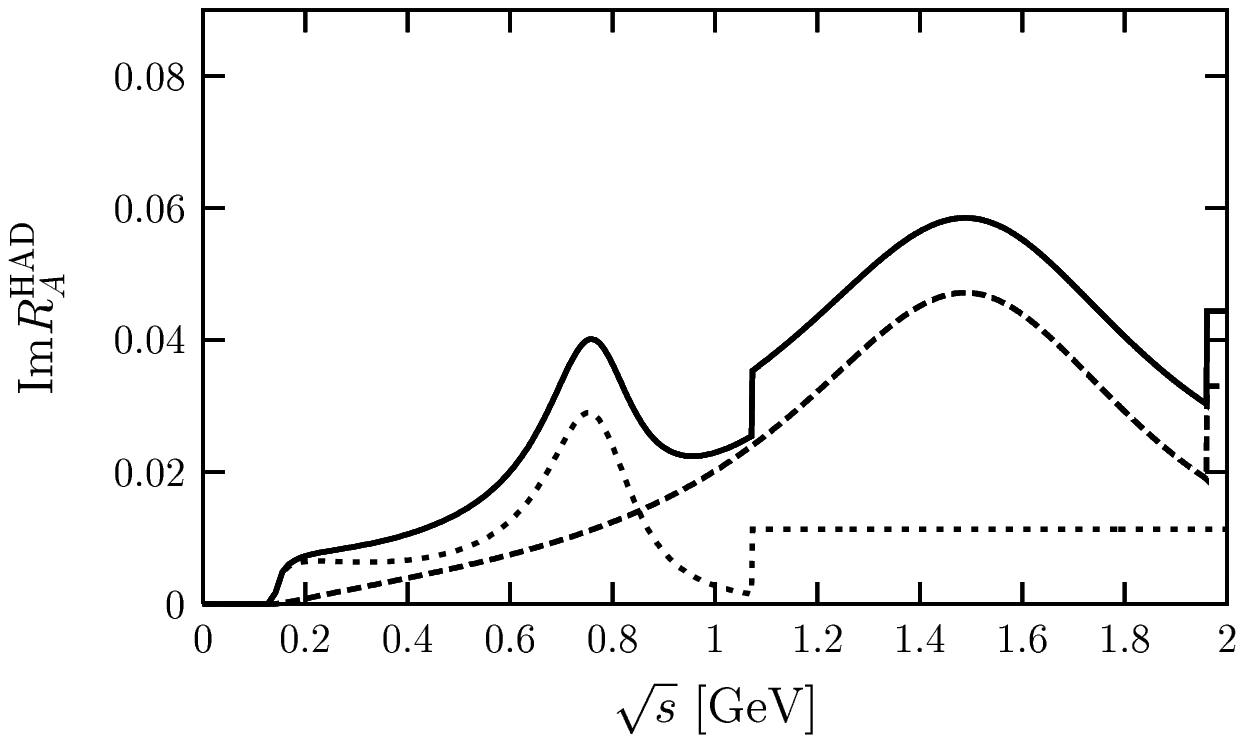}}
\caption{\label{fig:fullspeca1} 
Imaginary part of the full in-medium axial-vector correlator (full line) as a 
function of $\sqrt{s} = \sqrt{q^2}$. The contribution of the bare axial-vector (vector)
correlator is the given by the dashed (dotted) line. See main text for details.}
\end{figure}

As already mentioned we have found that there is a smaller in-medium shift of the 
mass-width band for the $\rho$ meson as compared to the old analysis which did
not include the mixing phenomena. At first glance this is a striking result. Suppose
that the old analysis is polluted by an additional $a_1$ peak, i.e.~a peak at large 
invariant masses. First of all, one might argue that this additional peak is suppressed
by the exponential function appearing in (\ref{eq:sumrule}). Therefore this peak should
not drastically modify the analysis for the $\rho$ meson as generally assumed 
for all details of the high-energy part. Note that
Figs.~\ref{fig:fullspecrho} and \ref{fig:fullspeca1} show that in general the position
of the $a_1$ peak is beyond the onset of the continuum of the $\rho$ branch, cf.~also
(\ref{eq:specval}). Let us ignore that
point for a moment and face a second apparent contradiction of our results to
naive expectations.
Consider the influence of the additional $a_1$ peak on the
determined mass (for a given width). Clearly the mass which we have attributed to the
$\rho$ meson in the old analysis would be a weighted average
of the true $\rho$ peak and the $a_1$ peak. Therefore the true $\rho$ peak should
be at smaller mass values as compared to the result of the old analysis. In the present
analysis, however, we have found just the opposite result. The solution to these
apparent contradictions is the appearance of the pion in the axial-vector branch.
It is not the high lying $a_1$ but the low lying pion which has dominantly pollutes 
the old
analysis. Since the pion is much lighter than the $\rho$ the former has caused a too
large shift of the mass-width band to the left in the old analysis. The $a_1$ peak
plays only a subdominant role as expected from modifications in the high energy regime.

After the decomposition of the full correlators into bare ones we have tacitly assumed 
that for the latter the one-peak ansatz for (the low energy part of) the spectral 
distribution is reasonable. In fact, also this assumption appears to be questionable
due to the coupling of the $\rho$-nucleon system to resonances 
(cf.~e.g.~\cite{peters,poust} and references therein). Especially the apparently
sizable coupling to the $D_{13}$ resonance $N^*(1520)$ might create an additional
peak at an invariant mass of roughly $580\,$MeV, i.e.~below the vacuum $\rho$ 
peak\footnote{Note that only resonances with an $s$-wave coupling to $N$-$\rho$ can 
contribute for the case at hand since we have considered $\rho$ mesons at rest with
respect to the nuclear medium. On account of the linear density approximation we have
neglected the Fermi motion of the nucleons.} --- provided that the resonance mass
does not change in the nuclear medium. Therefore e.g.~our result for the in-medium
mass-width band of the $\rho$ branch might still be polluted by the influence of
additional distinct peaks. We have refrained from including more than one peak in the 
ansatz for the spectral distribution since we would have been forced to introduce much 
more free parameters which cannot be determined from the sum rule. Further work is 
necessary to separate the different peaks. On the other hand, we have already seen
that the present sum rule analysis supports broad in-medium spectral functions. It
may appear that the baryonic resonances also melt in a nuclear environment 
(cf.~e.g.~\cite{moselerice} and references therein) and leave
no sign of distinct peaks in the mesonic channels. 

Concerning the included mixing effect of the vector and axial-vector channel the
amount of mixing is determined by the value for $\sigma^\pi_N$. As we have already 
mentioned even the sign of that quantity is still subject to discussion. The influence
of $\sigma^\pi_N$ is two-fold. First, it yields the amount of mixing on account of
(\ref{eq:mixing}). Here a change of $\sigma^\pi_N$ does not influence the sum rule 
analysis for the bare correlators but it does influence the final result for the
full correlators exemplified in Figs.~\ref{fig:fullspecrho} and \ref{fig:fullspeca1}.
Especially if $\sigma^\pi_N$ was negative one could no longer guarantee that 
the sign of the imaginary parts of the full correlators remains positive for all
values of the invariant mass $\sqrt{s}$. In this case the approximations which have
led to (\ref{eq:mixing}) should be revised. 

The second place where $\sigma^\pi_N$ appears
is the in-medium dependence of the quark condensates. Most importantly, it enters the
scaling assumption (\ref{eq:fourdensbare}) for the four-quark condensate via
(\ref{eq:splitnuclsig}). We have
made this assumption in lack of any better, more fundamental approach. One might
corroborate our choice by the expectation that factorization roughly works as long as 
there is no fundamental (symmetry) principle which is in contradiction to factorization.
In our case the interaction with the pions is determined by chiral symmetry. Here
factorization breaks down as proven for the case of finite temperature in 
\cite{eletsky}. Having split off the pions one might expect that the factorization 
assumption works for the interaction of the vector and axial-vector mesons with the 
bare nucleons. Nonetheless it is important to figure out how important the influence
of the four-quark condensate and especially of its in-medium change actually is:
At a typical value for the Borel mass of $M \approx 1\,$GeV we find the following
in-medium changes on the OPE side of the $\rho$ meson sum rule at nuclear saturation
density: The important changes
are induced by the twist-two condensate with dimension four, $2.7\cdot 10^{-4}$, and the
in-medium change of the four-quark condensate, $1.5\cdot 10^{-4}$. All other changes are
an order of magnitude smaller: 
$-3.6\cdot 10^{-5}$ from the in-medium change of the gluon condensate,
$1.3\cdot 10^{-5}$ from the in-medium change of the two-quark condensate,
$-2.7\cdot 10^{-5}$ from the twist-two condensate with dimension six.
Therefore even if the change of the four-quark condensate was completely neglected
roughly two thirds of the in-medium change would still persist. Thus, the results
obtained here would not drastically change. The vacuum contribution of the four-quark
condensate is $-4.8\cdot 10^{-4}$. If the four-quark condensate vanished completely 
at nuclear saturation density the in-medium change would be twice as large as the one
used in our analysis. One should realize that the previous considerations are somewhat
oversimplified as the dependence on the Borel mass is different for the OPE contributions
from the dimension-four condensates as compared to the ones with dimension six.
In total, however, we do not expect qualitative changes of the picture presented
here even if the in-medium behavior of the four-quark condensate was completely 
different.

In spite of the large uncertainties connected with the QCD sum rule method at finite
baryon density we regard it as a useful complement to purely hadronic approaches since
it connects in a unique way hadronic with quark-gluon degrees of freedom. 
We hope that the mentioned uncertainties can be removed step by step in the future.

\acknowledgments The author wants to thank Ulrich Mosel for discussions and continuous
support.

\end{document}